\documentclass[fleqn,usenatbib]{mnras}

\usepackage{newtxtext,newtxmath}

\usepackage[T1]{fontenc}

\DeclareRobustCommand{\VAN}[3]{#2}
\let\VANthebibliography\thebibliography
\def\thebibliography{\DeclareRobustCommand{\VAN}[3]{##3}\VANthebibliography}

\usepackage{graphicx}	
\usepackage{amsmath}
\usepackage{subcaption}
\usepackage{fix-cm}

\title{Kinematic Study of Molecular Gas in Cometary Globule - LBN 437}

\author[Aardra et al.]{
S. Aardra$^{1,2}$\thanks{E-mail: aardra.s@res.christuniversity.in},
Namitha Issac$^{2,3}$,
Archana Soam$^{2}$,
Blesson Mathew$^{1}$\thanks{E-mail: blesson.mathew@christuniversity.in},
Chang Won Lee $^{4,5}$
\\
$^{1}$Department of Physics and Electronics, CHRIST (Deemed to be University), Bangalore 560029, India\\
$^{2}$Indian Institute of Astrophysics (IIA), Koramangala, Bangalore 560034, India\\
$^{3}$Shanghai Astronomical Observatory, Chinese Academy of Sciences, 80 Nandan Road, Shanghai 200030, People’s Republic of China\\
$^{4}$Korea Astronomy and Space Science Institute (KASI), 776 Daedeokdae-ro, Yuseong-gu, Daejeon 305–348, Republic of Korea\\
$^{5}$University of Science \& Technology, 176 Gajeong-dong, Yuseong-gu, Daejeon, Republic of Korea\\}

\date{Accepted XXX. Received YYY; in original form ZZZ}

\pubyear{\the\year{}}

\begin{document}
\label{firstpage}
\pagerange{\pageref{firstpage}--\pageref{lastpage}}
\maketitle

\begin{abstract}
Bright-rimmed, cometary-shaped star-forming globules, associated with H{\sc ii} regions, are remnants of compressed molecular shells exposed to ultraviolet radiation from central OB-type stars. The interplay between dense molecular gas and ionizing radiation, analysed through gas kinematics, provides significant insights into the nature and dynamic evolution of these globules. This study presents the results of a kinematic analysis of the cometary globule, Lynds$^{\prime}$ Bright Nebula (LBN) 437, focusing on the first rotational transition of $^{12}$CO and C$^{18}$O molecular lines observed using the Taeduk Radio Astronomy Observatory (TRAO). The averaged $^{12}$CO spectrum shows a slightly skewed profile, suggesting the possibility of a contracting cloud. The molecular gas kinematics reveals signatures of infalling gas in the cometary head of LBN 437, indicating the initial stages of star formation. The mean infall velocity and mass infall rate towards the cometary head of LBN 437 are 0.25 km\,s$^{-1}$ and 5.08 $\times$ 10$^{-4}$ M$_{\odot }$\,yr$^{-1}$ respectively, which align well with the previous studies on intermediate or high-mass star formation.
\end{abstract}




\section{Introduction}
\begin{figure}
	\includegraphics[width=1.\columnwidth]{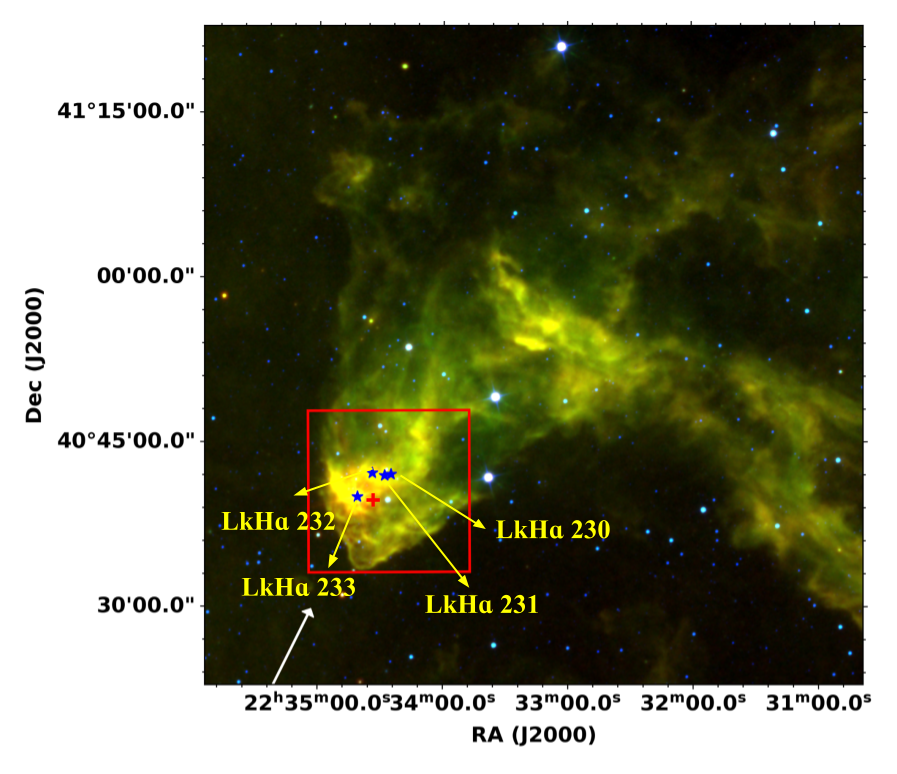}
    \caption{The WISE color-composite image of LBN 437 with the WISE 4.6 $\mu$m band assigned in blue, the 12 $\mu$m band in green, and the 22 $\mu$m band in red. The white arrow indicates the direction of ionizing radiation from 10 Lac. The square in red highlights the region selected for the analysis of molecular line emission. The plus symbol in red denotes the molecular concentration, Condensation A, described by \citet{olano1994molecular}. The asterisks in blue depict the four H$\alpha$ emission line sources in the cometary head of LBN 437.}
    \label{fig:CompositeImage}
\end{figure}

Massive, OB-type stars play a pivotal role in the evolution of their parental molecular cloud due to their substantial radiative, mechanical, and chemical feedback into the circumstellar environment via stellar winds and ionizing radiation \citep{2012MNRAS.427..625W}.  Despite their importance, the formation of these stars remains poorly understood, both theoretically and observationally. Core collapse and competitive accretion are the two competing models proposed to explain the formation of massive stars. In the core accretion model, the mass is accumulated prior to the formation of the massive star and the star formation is similar to low-mass star formation including formation via monolithic collapse, higher disc accretion, and outflow \citep{2002ApJ...569..846Y, 2003ApJ...585..850M, Tan_2014, 2018ARA&A..56...41M}. In contrast, the competitive accretion model proposes that the mass is gathered throughout the star formation process, as the gas is drawn toward the center of the cluster \citep{2001MNRAS.323..785B, 2006MNRAS.370..488B, Tan_2014, 2018ARA&A..56...41M}.\\ Once formed, these stars act as the principal source of ultraviolet radiation, which compresses and shapes the exposed material, thereby influencing the dynamics and chemistry of the surrounding environment. The high-energy photons emitted at the rates of  10$^{47}$\textendash\,10$^{50}$ s$^{-1}$ photoionize their environment, creating an over-pressured ionized bubble \citep{Saha_2022}. This bubble subsequently expands, forming a thin, unstable shell at the boundary of the H{\sc ii} region. The compressed molecular shell eventually leads to the formation of highly irregular and peculiar structures such as pillars, elephant trunks, and cometary globules (CGs), which protrude into the ionized gas.\\
CGs are isolated, relatively small molecular clouds detached from the parental molecular clouds with a cometary or a tear-shaped morphology. CGs are found in the vicinity of OB associations in the H{\sc ii} regions \citep{Hawarden, Sandqvist, Sugitani, Maheswar}. These interstellar clouds are characterized by a compact, bright-rimmed head and a faintly luminous, diffuse tail extending away from the early-type star. CGs are a subgroup of Bok globules with a size of $\sim$ 0.1 \textendash\,1.0 pc \citep{Hawarden,Zealey,2010RMxAA..46...67K} and mass ranging from 10 \textendash\,100 M$_{\odot }$ \citep{lefloch1994cometary,Haikala}. They also exhibit high number densities of 10$^{4}$ \textendash\,10$^{5}$ cm$^{-3}$  \citep{Vilas-Boas,bourke1995studies,Haikala}. Numerous studies also identify CGs as potential sites of low-mass star formation \citep[e.g.][]{Williams, Brand,reipurth1983star,mookerjea2009star, Makela}.\\
The formation and evolution of CGs have been attributed to several mechanisms, including collect and collapse \citep{1977ApJ...214..725E}, radiation-driven implosion \citep{sandford1982radiation, bertoldi1989photoevaporation, lefloch1994cometary}, shadowing \citep{2006RMxAA..42..203C, 2010MNRAS.403..714M}, and collapse due to shell curvature \citep{2012A&A...546A..33T}. However, these mechanisms, which depend on idealized conditions, oversimplify the complexities involved by neglecting the interplay of physical processes such as magnetic fields, turbulence, and protostellar feedback. Therefore, the impact of these mechanisms on cloud dynamics and star formation is not well understood. Using three-dimensional numerical simulations, \citet{2012A&A...546A..33T} investigated the interplay between ionizing radiation and turbulent molecular cloud and the subsequent formation of the CGs.  Studies also reveal that the formation of CGs is affected by the initial morphology and the clumpiness of the molecular cloud \citep[e.g.][]{walch2013clumps}. But the origin of CGs is still uncertain, whether they are the result of pre-existing dense structures in the surrounding interstellar medium advected through the ionization front \citep{reipurth1983star} or flow instabilities \citep{1996ApJ...469..171G,williams1999shadowing}.\\
The cometary globule Lynds$^{\prime}$ Bright Nebula (LBN) 437 (\textit{l} $=96^{\circ}$, \textit{b} $=-15^{\circ}$), is located at the edge of an elongated molecular cloud complex Kh149 \citep{khavtassi1960atlas} at the boundary of H{\sc ii} region, S126 \citep{sharpless}. The formation of the cometary head of LBN 437 and the triggered star formation occurring in the cloud is attributed to the interaction of the cloud with the ionizing radiation from the Lac OB1 members \citep{olano1994molecular}. The brightest of the Lac OB1 members is the star 10 Lac with a spectral type O9V, located south-east of LBN 437, at an angular distance of $\sim$\,1.8$^{\circ}$. The nuclear region of LBN 437 also coincides with a reflection nebula, DG187 \citep{Dorschner}. LBN 437 lacks a prominent tail and resembles a comma-like morphology. Considering the spatial and kinematic coincidence of the cloud with the Lac OB1 members, \citet{olano1994molecular} estimated the distance to LBN 437 as 460 pc.  Later, using a near-infrared photometric method,  \citet{soam2013magnetic} determined a distance of 360\,$\pm$\,65 pc to LBN 347.\\
\citet{olano1994molecular} resolved the southern part of Kh149 into two condensation nuclei, namely Condensation A and B in $^{12}$CO and $^{13}$CO molecular line observations. Their observations of $^{12}$CO line profiles provided evidence of blue-skewness, with the blue-shifted peak more intense than the red-shifted peak. However, the underlying kinematic processes driving these features remained unexamined. The molecular concentration named Condensation A in the cometary head is associated with an optical reflection nebula and houses a group of H$\alpha$ emission line sources, namely LkH$\alpha$ 230, LkH$\alpha$ 231, LkH$\alpha$ 232, and LkH$\alpha$ 233. It contains a cold, elliptical, dense core traced by NH$_{3}$ emission. LkH$\alpha$ 233 (or V375 Lac)  in the Condensation A, identified as a Herbig A4e star \citep{hernandez2004spectral}, shows emission lines such as H$\alpha$, [OI] 6300~\AA, [SII] 6717~\AA \, in its spectrum \citep{leechen}. The formation of these young stars might be triggered by radiation-driven implosion, in which UV radiation from a luminous star such as 10 Lac causes the evaporation and compression of the molecular cloud \citep{olano1994molecular}. The less massive Condensation B near the end of the cometary tail is not associated with any optical stars.  Figure~\ref{fig:CompositeImage} shows the WISE color-composite image of the LBN 437 using 4.6 $\mu$m~(blue), 12 $\mu$m~(green), and 22 $\mu$m~(red) images of size 1$^{\circ}$ $\times$ 1$^{\circ}$. Assuming 10 Lac to be the ionizing source, we indicate the direction of ionizing radiation from 10 Lac towards LBN 437 using a white arrow. We also depict the four H$\alpha$ emission line sources and Condensation A in the cometary head of LBN 437.
\citet{soam2013magnetic} studied the magnetic field geometry at the periphery of LBN 437 using the R-band polarimetry and found that the magnetic field lines in the globule follow the curvature of the globule head. But before this scenario, with no effects of the ionizing radiation, the initial magnetic field lines were found to be perpendicular to the direction of the ionizing radiation. This change in orientation is attributed to the drag that the magnetic field lines experienced due to the ionizing radiation from the same source responsible for the cometary morphology of LBN 437.
\\
Previous studies by \citet{olano1994molecular} and \citet{soam2013magnetic} have identified that the ionizing radiation from Lac OB1 members shapes the cometary morphology of LBN 437 and can possibly trigger star formation in the cometary head. This can, in turn, affect the gas motion in LBN 437. Although \citet{olano1994molecular} investigated the region using multiple molecular line tracers, their study did not examine the kinematic structure of the cloud. Hence, a detailed kinematic analysis is crucial to understand the dynamics of the molecular gas and the physical processes driving star formation in this region. The molecular line observations of CO are critical in understanding the dynamics of the cloud as they provide direct evidence of gravitational collapse and aid in understanding the early stages of star formation \citep{Schneider_2015}. They also explain the influence of UV radiation and stellar feedback from the nearby massive stars on the gas dynamics. The rotational transitions of CO in the millimeter wavelengths are primary tracers of molecular gas in the clouds \citep{Pe_aloza_2016}. $^{12}$CO and C$^{18}$O molecular lines trace various velocity components within the CGs, revealing features such as outflow, turbulence, and rotation. In particular, the lower-J rotational transitions of CO reveal the molecular outflow occurring in the star-forming regions \citep{doi:10.1146/annurev-astro-081915-023341, Cortes-Rangel_2020}. A robust understanding of these physical mechanisms active in the CGs will provide useful insight into cloud dynamics and their impact on the star formation process.\\
In this study, we perform the kinematic investigation of molecular gas associated with the cometary head of LBN 437 using molecular data extracted from the Taeduk Radio Astronomy Observatory (TRAO). A detailed study of the molecular gas associated with the region and the critical inspection of the molecular line profiles correlate the gas kinematics with the dynamic processes occurring in the cloud. To the best of our knowledge, this is the first study dedicated to probing the molecular gas kinematics of LBN 437, thereby offering a new perspective on its star formation dynamics. The complementary optical depths of C$^{18}$O (optically thin) and $^{12}$CO (optically thick) probes distinct cloud components; C$^{18}$O reveals the denser, embedded cores inside the cloud, whilst $^{12}$CO mainly traces the surface layers \citep{2009A&A...497..789U}. The combination of $^{12}$CO and C$^{18}$O used in this study provides a reliable first-order approximation of the large-scale gas dynamics, even though the addition of other molecular tracers would provide a more thorough characterisation of the cloud. The paper has been organized in the following manner: Section~\ref{sec:Observational and Archival Data} describes the observational and archival data used in the study. Section~\ref{sec:Results} elaborates on the CO emission observed towards the cometary head of LBN 437 and the velocity structure of the molecular gas. Section~\ref{sec:Kinematic Signatures} discusses the kinematic signatures of molecular gas, such as the infall motion. Section~\ref{sec:Summary} compiles the summary of the study.
\section{Observational and Archival Data}
\label{sec:Observational and Archival Data}
\subsection{Observational Data}
To understand the kinematics of molecular gas in LBN 437, an On-The-Fly (OTF) mapping observation of the cometary head of LBN 437 was performed in $^{12}$CO\,(1\textendash\,0) and C$^{18}$O (1\textendash\,0) molecular lines simultaneously. This was carried out using the SEcond QUabbin Observatory Imaging Array (SEQUOIA) at the Taeduk Radio Astronomical Observatory (TRAO), which is equipped with high-performing 16-pixel MMIC preamplifiers in a 4\,$\times$\,4 array. Located at the Korea Astronomy and Space Science Institute (KASI) in Daejeon, South Korea, TRAO is a 13.7-m Radio Telescope with a single-horn receiver system, operating in a frequency range of 86\,\textendash 115 GHz. The system temperature ranges from 150 K (86$-$110 GHz) to 450 K (115 GHz; $^{12}$CO). Since the optical system offers two sidebands, it is possible to observe the $^{12}$CO\,(1\textendash\,0) and C$^{18}$O (1\textendash\,0) molecular lines simultaneously. \\
The primary beam efficiency and beam size Half Power Beam Width (HPBW) at 115 GHz were found to be 54\,$\pm$\,2, and around 44 arcsec,  respectively. Orion A in the SiO line was chosen as the source to point and focus the telescope. The pointing of the telescope was as accurate as 5\,\textendash 7 arcsec. The velocity resolution achieved was found to be $\sim$ 0.1 km\,s$^{-1}$. The sky signals were subtracted in the position switch mode. During the observation, the system temperature was between 500 and 650 K. The integration time of the observation was $\sim$\,180 min to achieve an rms of 0.3 K in $T_{\rm A}$$^{\ast}$. The data reduction was done using the CLASS software of the GILDAS package.     

\subsection{Archival Data}
\subsubsection{Planck Data}
\textit{Planck} provided the first all-sky map of the polarized thermal emission from the galactic dust at submillimeter wavelengths \citep{2016A&A...586A.138P}. The High Frequency Instrument (HFI) optically coupled to the \textit{Planck} telescope through cold optics at 4 K, 1.6 K, and 0.1 K, is designed around 52 bolometers cooled to 0.1 K and observe the sky over a frequency range of 100 to 857 GHz \citep{2014A&A...571A...8P, 2016A&A...594A...8P}. In this study, we used the \textit{Planck} emission map at 353 GHz with an angular resolution of 5 arcmin to examine the cold dust emission in the cometary head of LBN 437. We also used the \textit{Planck}-HFI brightness emission map at 857 GHz with an angular resolution of 5 arcmin to estimate the mass of the cometary head of LBN 437.
\subsubsection{Wide-field Infrared Survey Explorer Data}
The Wide-field Infrared Survey Explorer (WISE; \citealp{Wright_2010}) provided a digital image atlas of the entire sky in four mid-infrared bands at 3.4 $\mu$m, 4.6 $\mu$m, 12 $\mu$m, and 22 $\mu$m ( W1, W2, W3, and W4, respectively). The angular resolution achieved was 6.1 arcsec, 6.4 arcsec, 6.5 arcsec, and 12.0 arcsec, respectively. We used the WISE  4.6 $\mu$m, 12 $\mu$m, and 22 $\mu$m images of LBN 437 to understand the properties of the warm dust emission associated with the cloud.  

\section{Results}
\label{sec:Results}
\subsection{Molecular emission in LBN 437}
\label{Molecular emission in LBN 437}
\begin{figure}
\centering
\begin{subfigure}{\columnwidth}
    \includegraphics[width=\textwidth]{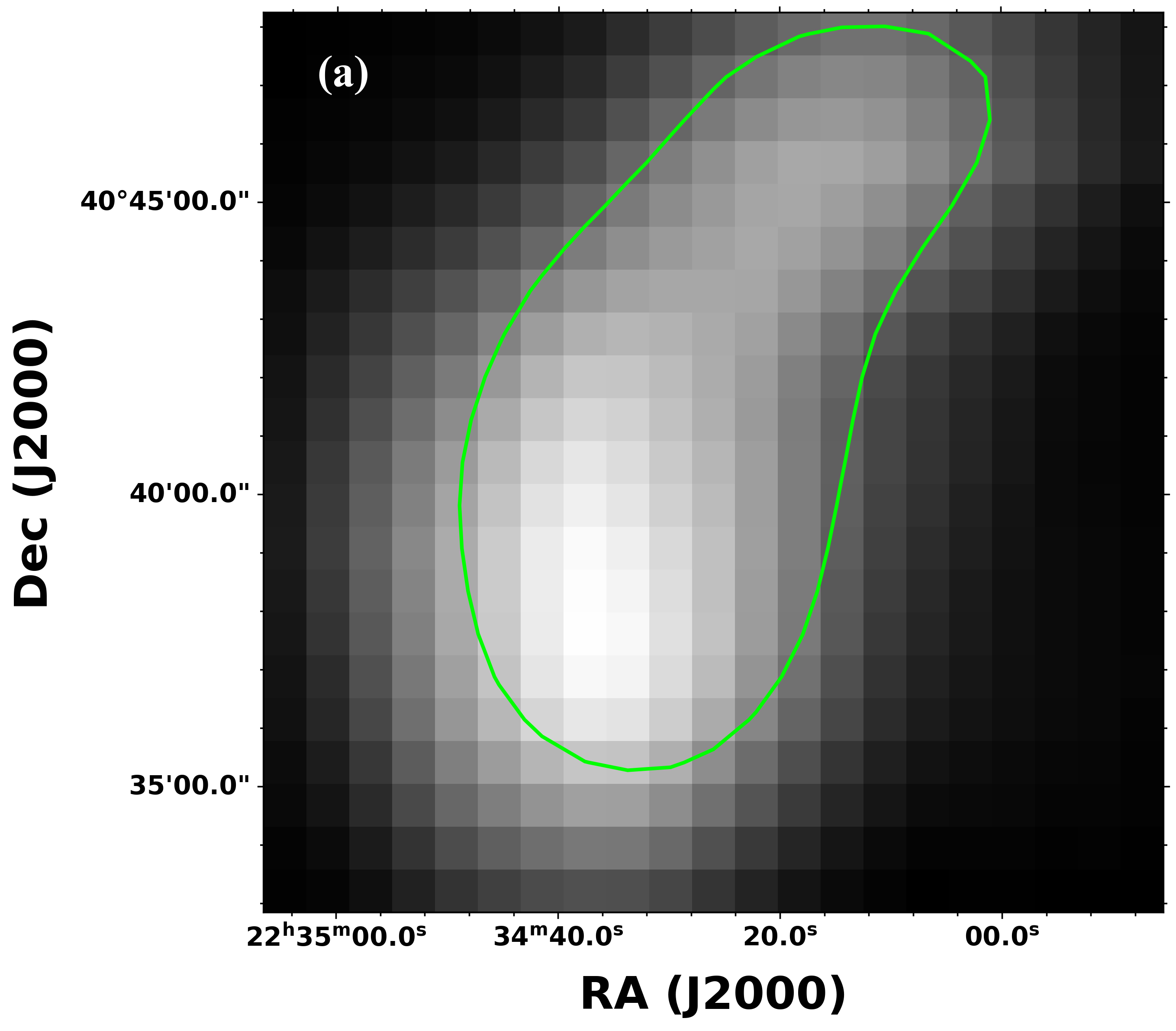}
\end{subfigure}
\hfill
\begin{subfigure}{1.1\columnwidth}
    \includegraphics[width=\textwidth]{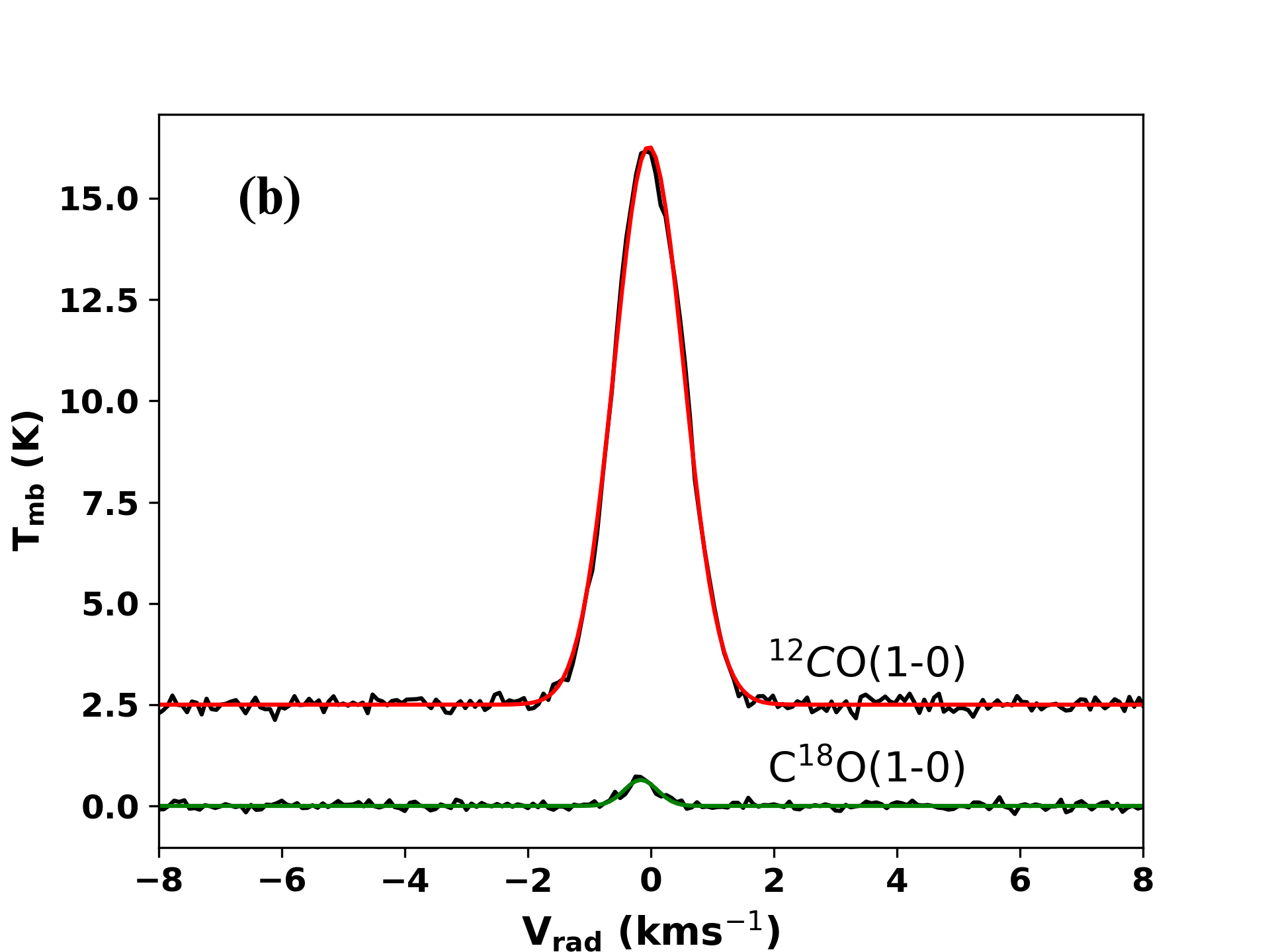}
\label{fig:figures}
\end{subfigure}
\caption{(a) Channel map of the $^{12}$CO\,(1\textendash\,0) emission covering a velocity range of -0.238 to -0.159 km\,s$^{-1}$. The green contour represents the half-maximum contour over which the $^{12}$CO\,(1\textendash\,0) and C$^{18}$O (1\textendash\,0) spectra are averaged. (b) The spectral line profiles of $^{12}$CO\,(1\textendash\,0) and C$^{18}$O (1\textendash\,0) emission averaged over the half-maximum contour and fitted with a single gaussian. The fit to the $^{12}$CO spectrum is depicted in red and C$^{18}$O in green.
}
\label{fig:Average Spectra}
\end{figure}

The CO molecular line emission sheds light on the evolutionary stage of a molecular cloud, its physical properties, gas properties, and kinematics \citep{2013ApJ...775L...2L, 2013A&A...556A.105O, Pe_aloza_2016, 2024MNRAS.528.2199R}. The diverse molecular line profiles help understand the kinematics of expansion, outflows, infall, and rotation inside the molecular cloud. We analysed the first rotational transition (J=1\textendash\,0) of $^{12}$CO and C$^{18}$O molecular line emission towards the cometary head of LBN 437. Since the first rotational transitions of the CO molecule are indicators of outflow activity, these are excellent for studying the star-forming regions \citep{2001ApJ...552L.167Z,2002A&A...383..892B}. The optically thick $^{12}$CO molecular line is an effective tracer of the kinematic and spatial extent of the outflow, whereas the optically thin C$^{18}$O line is a tracer of high-density cloud cores \citep{2015MNRAS.453.3245L}.\\ We present the averaged $^{12}$CO and C$^{18}$O spectra over the half-maximum contour in the $^{12}$CO intensity map. We fit a single gaussian profile to the C$^{18}$O (1\textendash\,0) and $^{12}$CO\,(1\textendash\,0) molecular emission to derive the main beam temperature ($T_{\rm mb}$), velocity width ($\Delta V$) and the peak velocity ($V_{\rm peak}$). The channel map with the half-maximum contour is shown in Figure~\ref{fig:Average Spectra}a, and the fitted profiles are shown in Figure~\ref{fig:Average Spectra}b. The retrieved parameters are listed in Table~\ref{tab:Parameters}. We observe a blue-skewness in the averaged $^{12}$CO\,(1\textendash\,0) spectrum, indicating that the cloud is undergoing contraction, a characteristic often associated with regions of active star formation. The line asymmetry suggests that the molecular gas in the cometary head is collapsing.  Such features are often observed in star-forming regions, where spectral line asymmetries have been interpreted as evidence of gravitational collapse \citep{1977ApJ...214L..73L, 1993ApJ...404..232Z, 1996ApJ...465L.133M, 2013A&A...555A.112P, 2019ApJ...870....5J}.
 To obtain the main beam temperature ($T_{\rm mb}$), we applied the beam correction to the antenna temperature ($T_{\rm A}$$^{\ast}$) using the formula $T_{\rm mb}$ = $T_{\rm A}$$^{\ast}$ $/$ $\eta_{\rm mb}$, where $\eta_{\rm mb}$ is the main-beam efficiency \citep{2014ApJ...786..140R}. We considered $\eta_{\rm mb}$ for TRAO as 0.54 following \citet{2018ApJS..234...28L}. The peak velocity obtained from the gaussian fit to C$^{18}$O (1\textendash\,0) line provides the local standard of rest velocity ($V_{\rm LSR}$). We found the $V_{\rm LSR}$ of the region to be close to 0.0 km\,s$^{-1}$.

\begin{table}
	\centering
	\caption{Peak velocities, velocity widths, and peak fluxes of the  $^{12}$CO\,(1\textendash\,0) and C$^{18}$O (1\textendash\,0) spectra averaged over the half-maximum contour in the $^{12}$CO intensity map (shown in Figure~\ref{fig:Average Spectra}).} 
	\label{tab:Parameters}
	\begin{tabular}{lccccr} 
		\hline
		Transition & $V_{\rm peak}$ & $\Delta V$ & $T_{\rm mb}$\\
		   & km\,s$^{-1}$ & km\,s$^{-1}$ & K\\
            \hline
            $^{12}$CO\,(1\textendash\,0) & -0.033 $\pm$ 0.002 & 0.564 $\pm$ 0.002 & 13.78 $\pm$ 0.05 \\
            C$^{18}$O (1\textendash\,0) & -0.163 $\pm$ 0.018 & 0.270 $\pm$ 0.018 & 0.643 $\pm$ 0.037\\
		\hline
	\end{tabular}
\end{table}

\subsection{Continuum emission in LBN 437}
\begin{figure}
	\includegraphics[width=1.03\columnwidth]{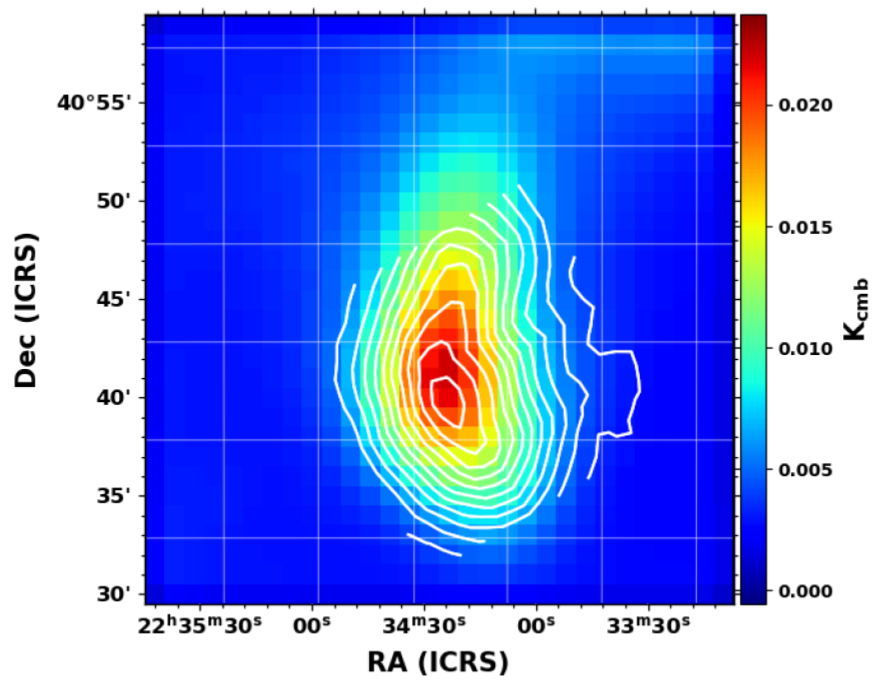}
    \caption{The zeroth moment map  (-20 km\,s$^{-1}$ to 20 km\,s$^{-1}$)  of $^{12}$CO\,(1\textendash\,0) overlaid on the 353 GHz \textit{Planck} map. The contours start from 3$\sigma$ level and increases in the steps of 8$\sigma$ where $\sigma$ = 0.167 K~m~s$^{-1}$. 
}
    \label{fig:Planckmap}
\end{figure}

CGs abundant in cold dust serve as direct indicators of star formation activity in these regions. Unlike molecular gas, dust in CGs responds differently to external radiation. Dust grains absorb high-energy photons and re-emit in infrared, heating the surrounding gas. However, the ionizing radiation causes the molecular gas at the periphery of the cloud to undergo photoevaporation \citep{2021RAA....21...87S}.  Examining the spatial correlation between dust emission and molecular gas distribution within CGs is key to understanding the impact of ionizing radiation in shaping the internal structures of CGs, which can trigger star formation or dissipate the cloud \citep{2008Ap&SS.315..215M, 2011MNRAS.415.1202C, Saha_2022,2022ApJ...928...17S}.
The \textit{Planck} 353 GHz map is pivotal in this analysis to understand the distribution of cold dust and its spatial correlation with the molecular gas emission in the cometary head of LBN 437. The 353 GHz frequency, corresponding to a wavelength of 850\,$\mu$m, falls within the far-infrared to submillimeter range, where dust emission is predominant, particularly from the cold dust grains. This makes the \textit{Planck} 353 GHz emission map sensitive to the emission from the cold dust grains. Figure~\ref{fig:Planckmap} illustrates the 0.5$^{\circ}$ $\times$ 0.5$^{\circ}$ \textit{Planck} map at 353 GHz overlaid with the zeroth moment map of $^{12}$CO emission. The contours of the $^{12}$CO zeroth moment map trace the same regions where the emission from the cold dust is observed, demonstrating the spatial correlation between the distribution of molecular gas and cold dust emission in the cometary head of LBN 437. A strong correlation suggests that the cometary head retains enough material density, which can cause the region to collapse and subsequently form stars.

\subsection{Velocity Structure}
\begin{figure}
	\includegraphics[width=1.1\columnwidth]{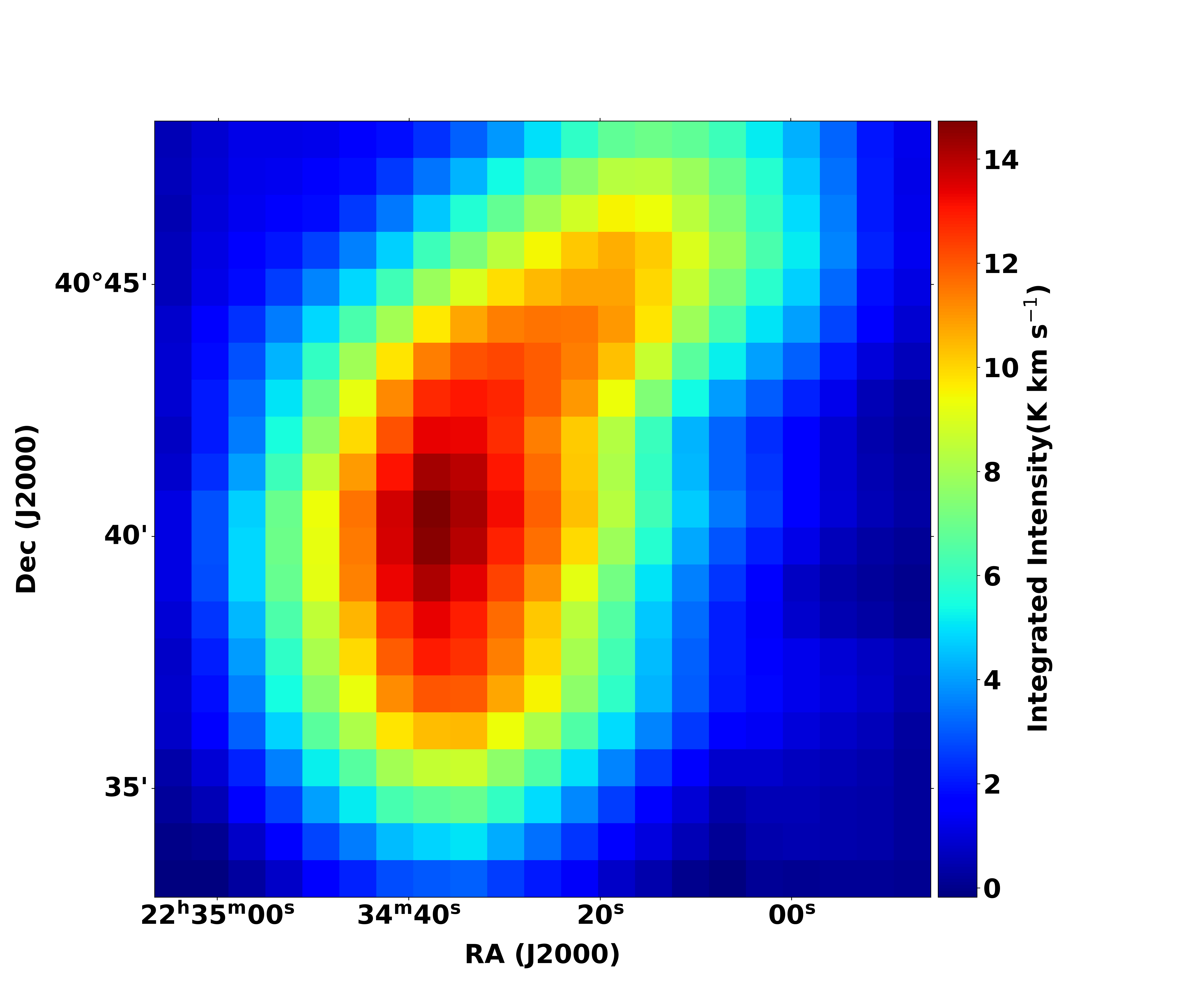}
    \caption{Zeroth moment map of $^{12}$CO\,(1\textendash\,0) emission in the cometary head of LBN 437. The specified velocity range is from -20 km\,s$^{-1}$ to 20 km\,s$^{-1}$.}
    \label{fig:Moment map}
\end{figure}

\begin{figure}
    \includegraphics[width=1.01\columnwidth]{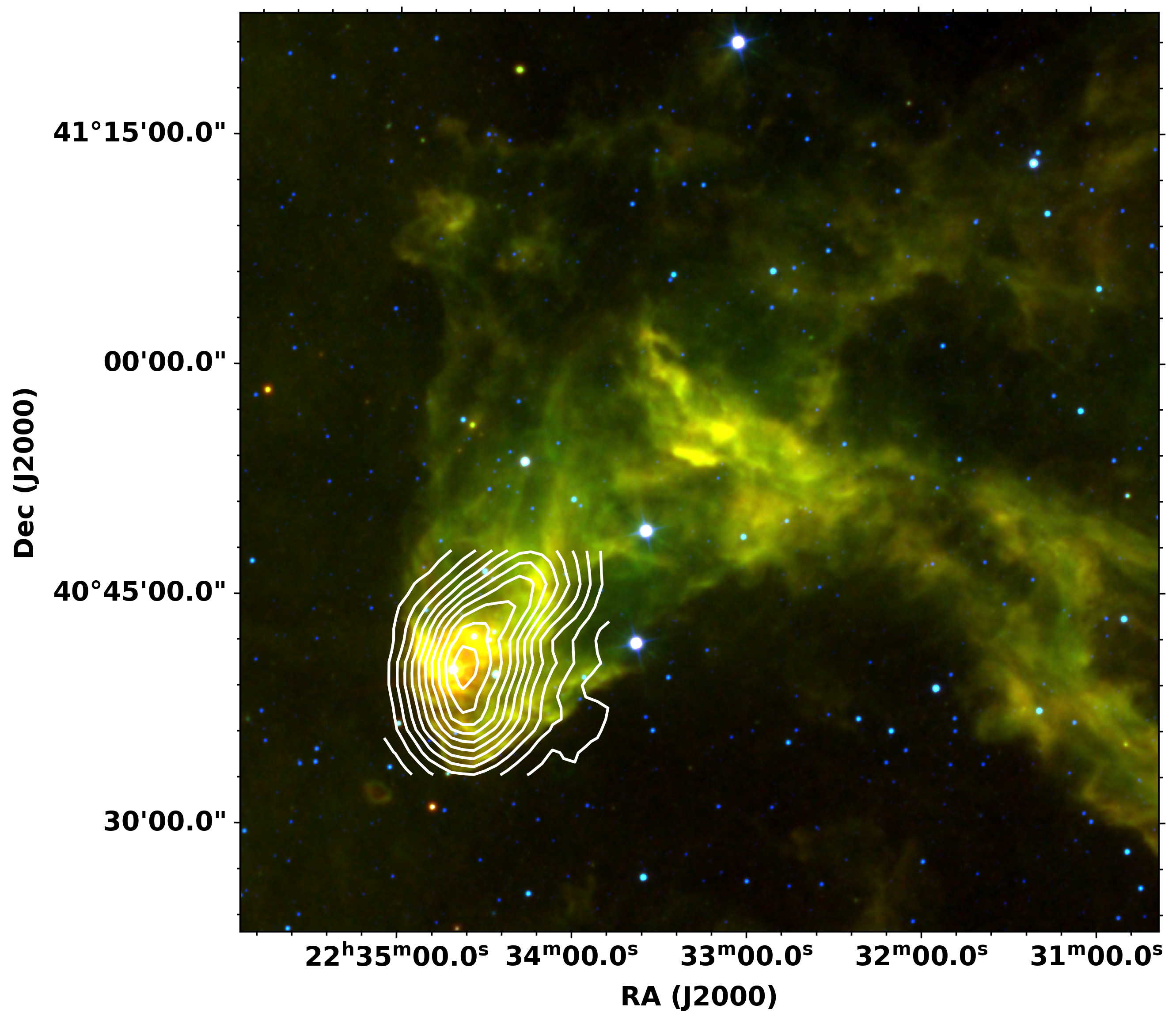}
    \caption{Zeroth moment map (-20 km\,s$^{-1}$ to 20 km\,s$^{-1}$)  of $^{12}$CO\,(1\textendash\,0) overlaid on the WISE color-composite image of LBN 437. The contours start from 3$\sigma$ level and increases in the steps of 8$\sigma$ where $\sigma$ = 0.167 K\,m\,s$^{-1}$.}
    \label{fig:Momentmapoverlay}
\end{figure}

\begin{figure*}
	\includegraphics[width=18cm, height=18cm]{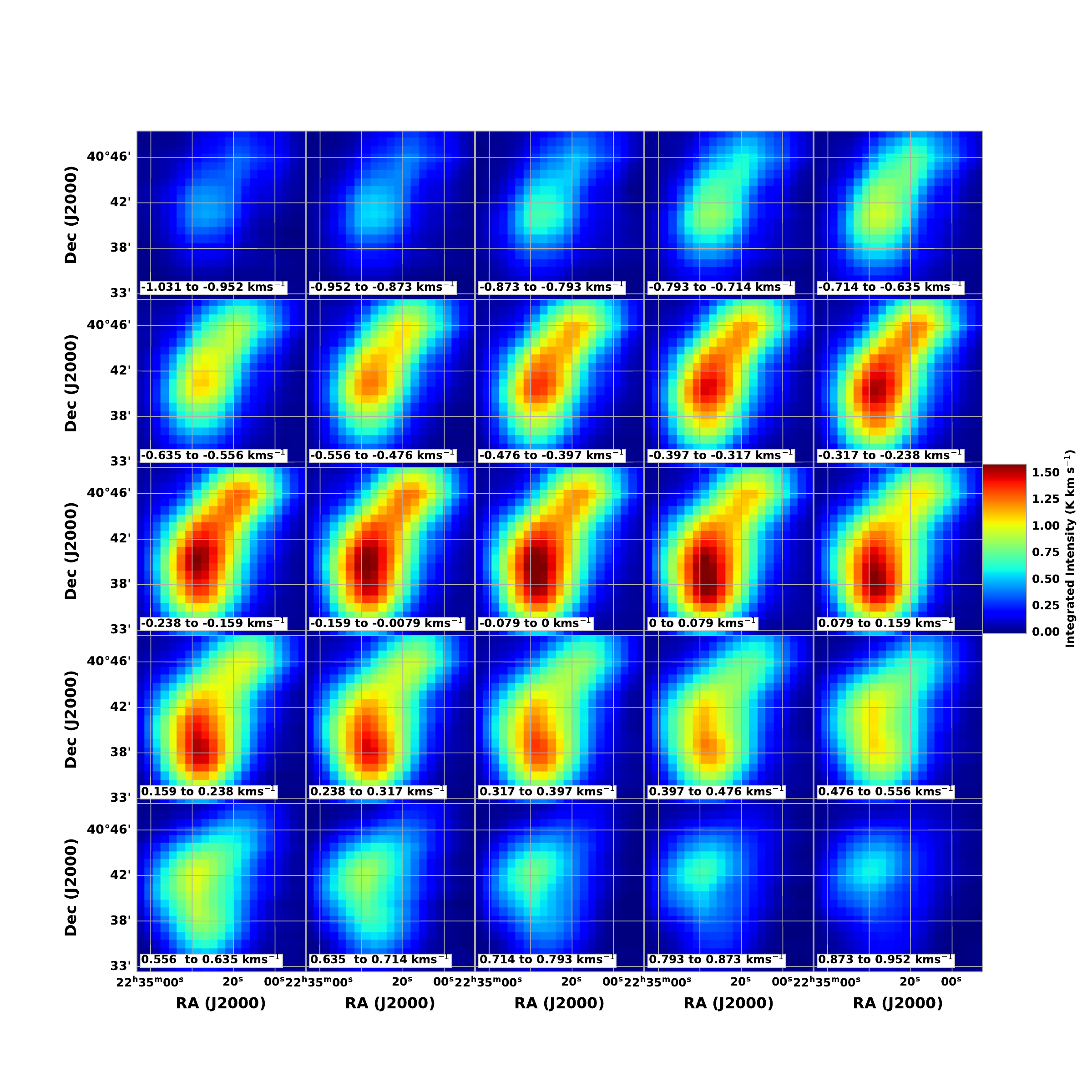}
    \caption{Channel maps of $^{12}$CO\,(1\textendash\,0) emission associated with the cometary head of LBN 437. The velocity width of each channel is 0.079 km\,s$^{-1}$.}
    \label{fig:Channelmap}
\end{figure*}

Performing gas kinematics is crucial for understanding the velocity structure of the cloud. Figure~\ref{fig:Moment map} presents the zeroth moment map of $^{12}$CO\,(1\textendash\,0). This map displays the intensity distribution of $^{12}$CO\,(1\textendash\,0) molecular species integrated over the designated velocity range from -20 to 20 km\,s$^{-1}$.
In Figure~\ref{fig:Momentmapoverlay}, we overlay the zeroth moment map of $^{12}$CO emission over the 1$^{\circ}$ $\times$ 1$^{\circ}$ WISE color-composite image of LBN 437 to analyse the relationship between the gas and warm dust emission in the cometary head of LBN 437. The moment zero contours trace the same regions exhibiting warm dust emission, indicating a coupled nature between the gas dynamics and warm dust emission in the observed regions. Understanding this coupling is essential to comprehend the physical processes occurring in the cometary head. The warm dust emission, traced by the WISE image, reveals regions heated either by external UV radiation from the Lac OB1 members or by feedback from young stars within the cloud. The molecular gas kinematics, derived from the $^{12}$CO emission, uncover the response of the molecular gas to these processes. The alignment of moment zero contours with warm dust emission suggests that the cometary head of LBN 437 is a dynamically active star-forming region. The cloud morphology appears to be strongly influenced by environmental factors such as UV radiation from the massive stars. This radiation likely induces feedback-driven compression, shaping the cometary structure and triggering star formation.\\
To illustrate the velocity structure of the cloud, we generate the channel map of $^{12}$CO emission associated with the region, as depicted in Figure~\ref{fig:Channelmap}. The velocity intervals specified in the channel map are determined by analyzing the $^{12}$CO TRAO data cube and detecting the $^{12}$CO  emission. The width of each channel is chosen as 0.079 km\,s$^{-1}$. The channel map reveals a distinct shift in the peak emission of $^{12}$CO across the different velocity ranges.
\section{Discussion}
\label{sec:Kinematic Signatures}
\subsection{Spectral Map of Molecular Emission}
\label{sec:Spectral Map of Molecular Emission}
\begin{figure*}
    \hspace{-0.7cm}
    \includegraphics[width=18cm,height =18cm]{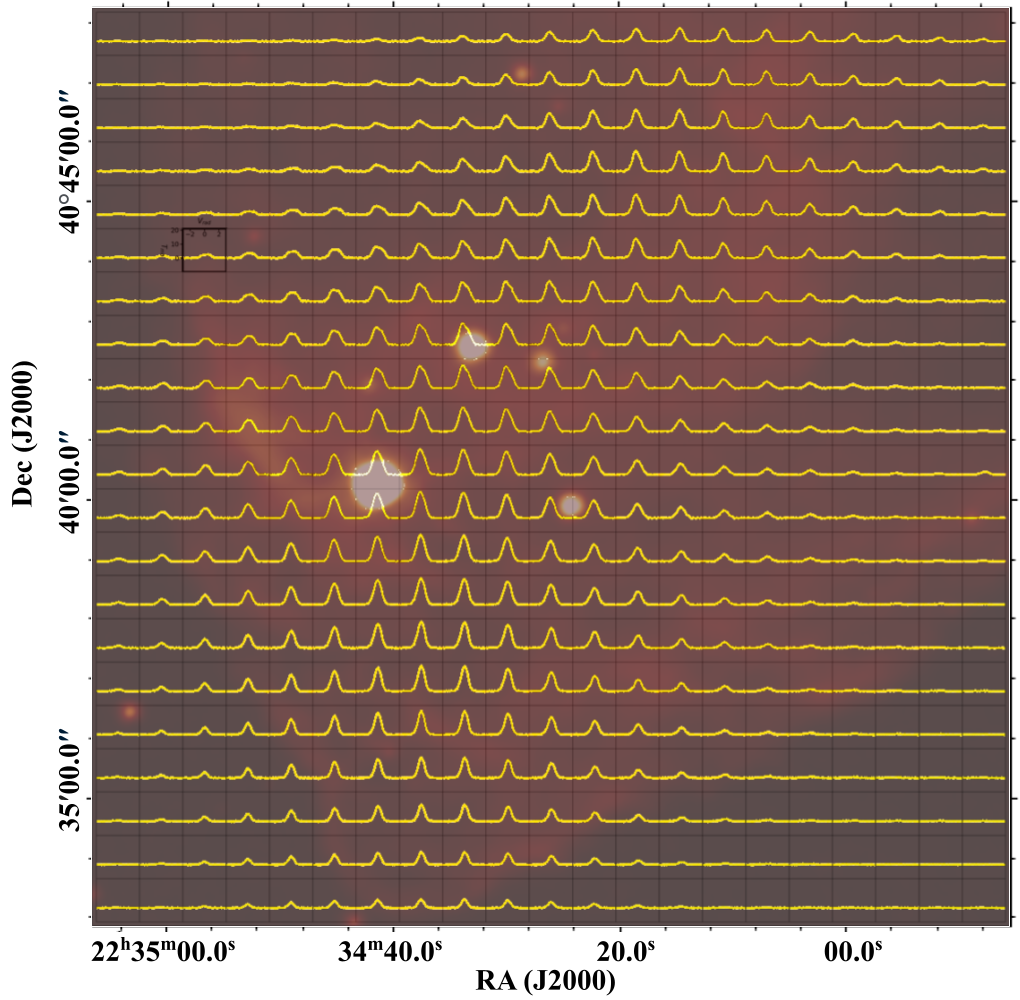}
    \caption{ Spectral grid map of $^{12}$CO\,(1\textendash\,0) emission overplotted on the  WISE 12 $\mu$m image of the cometary head of LBN 437. The line emission is binned into 21 $\times$ 21 grid, each of size 45 arcsec over which the $^{12}$CO\,(1\textendash\,0) spectrum is extracted. In the spectral grid, the radial velocity ($V_{\rm rad}$) on the X-axis ranges from -3 to 3 km\,s$^{-1}$, and the main beam temperature ($T_{\rm mb}$) on the Y-axis ranges from 0 to 20 K.}
    \label{fig:GridmapB}
\end{figure*}

A grid map of the molecular emission observed in the cometary head of LBN 437 facilitates an understanding of the gas motion within the cloud. We construct the grid map of $^{12}$CO molecular line profiles in the cometary head of LBN 437 by binning the line emission into a 21 $\times$ 21 grid, each of size 45 arcsec over which the spectra is extracted. We overlay the grid map on the 12$\mu$m WISE image of the cometary head of LBN 437 as shown in Figure~\ref{fig:GridmapB}. The spectral map of $^{12}$CO molecular emission reveals asymmetric signatures in the $^{12}$CO line profiles. The asymmetric profiles are most significant in the vicinity of the LkH$\alpha$ sources seen as bright sources on the spectral map. The predominant blue-skewness indicates the occurrence of cloud contraction, consistent with scenarios of gravitation collapse. However, the $^{12}$CO line profiles are not blue-skewed uniformly in every bin.  There are also some localized regions where we observe nearly Gaussian profiles. Notably, the optically thick \(^{12}\)CO emission in the inner regions of the cometary head shows a dominant blue-skewness, reinforcing the evidence for infall motion and suggesting that the cometary head of LBN 437 is undergoing contraction.
\subsection{Kinematic Signatures of Infalling Gas} 
The infall motion of molecular gas, typically identified through the molecular line profiles and their asymmetries, supports the model of a gravitationally collapsing cloud \citep{Gao}. As discussed in Sections \ref{Molecular emission in LBN 437} and \ref{sec:Spectral Map of Molecular Emission}, the averaged $^{12}$CO spectrum and spectral map of $^{12}$CO molecular emission display asymmetric line profiles, notably, blue-skewness. We analyse the asymmetric line profiles of $^{12}$CO relative to the optically thin C$^{18}$O. To examine and quantify the infall motion of the molecular gas in the observed regions, we estimate the infall velocity and the mass infall rate towards the cometary head of LBN 437. 
\subsubsection{Blue-skewed line profiles of $^{12}$CO}
Optically thick lines are well-known tracers of gas motion along the line of sight \citep[e.g.][]{2005A&A...442..949F, Smith_2012,2013A&A...555A.112P}. Due to their sensitivity to excitation gradients, these lines are generally used to infer the dynamic processes occurring in the cloud. The appearance of "blue-skewness" in the line profiles of optically thick lines serves as a spectral signature of the infall motion \citep{1994ASPC...65..183Z,1997ApJ...489..719M,1999ApJ...526..788L,2005A&A...442..949F,2019ApJ...870....5J,10.1093/mnras/stab2801}. In the infall scenario with the excitation temperature increasing towards the center and a radial temperature gradient, the blue-shifted gas arises from the inner envelopes at a higher excitation temperature than the red-shifted gas emerging from the outer envelope \citep{1977ApJ...211..122S,1994ASPC...65..183Z,1999ARA&A..37..311E}. Hence, in a blue-skewed spectral line profile, the blue-shifted peak is observed to be more intense than the red-shifted peak. Searching for "blue-skewness" in the line profiles of optically thick spectral lines is a primary technique for identifying the gravitational collapse occurring in the cloud \citep{Gao,2013A&A...555A.112P,10.1093/mnras/stx2480,2019ApJ...870....5J}.
\subsubsection{Characterizing  Infall Motion}
\begin{figure}
	\includegraphics[width=\columnwidth]{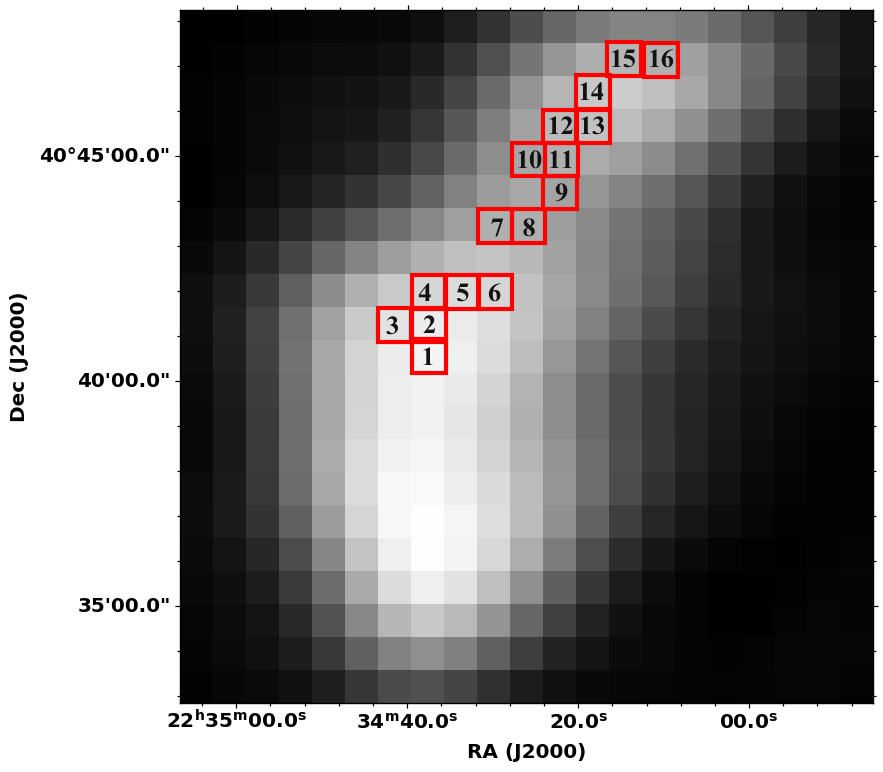}
    \caption{$^{12}$CO\,(1\textendash\,0) channel map of cometary head of LBN 437 with a velocity range of  -0.238 to -0.159 km\,s$^{-1}$. The red squares are the regions of size 45 arcsec over which the spectra of $^{12}$CO and C$^{18}$O  are averaged.
}
    \label{fig:regions}
\end{figure}

\begin{figure*}
    \centering
    \begin{subfigure}[t]{0.35\textwidth}
        \centering
        \includegraphics[height=1.85in]{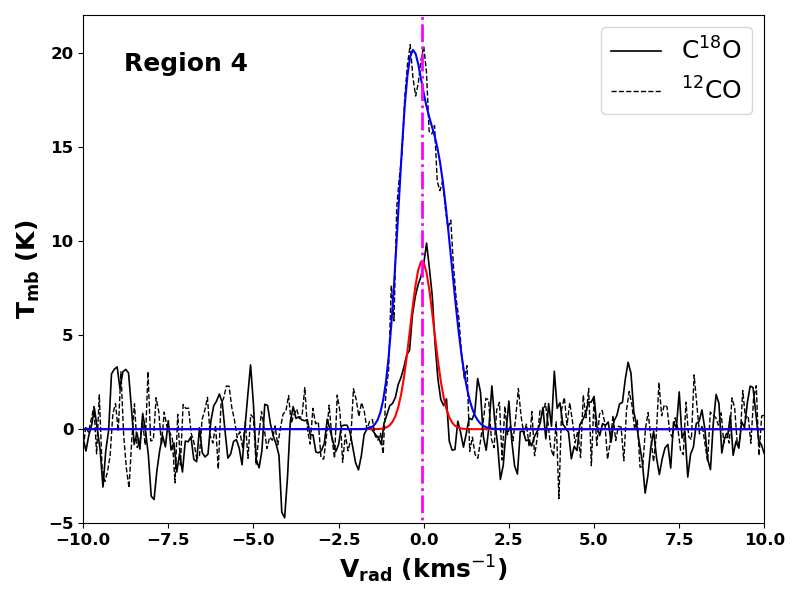}
    \end{subfigure}%
    \begin{subfigure}[t]{0.35\textwidth}
        \centering
        \includegraphics[height=1.85in]{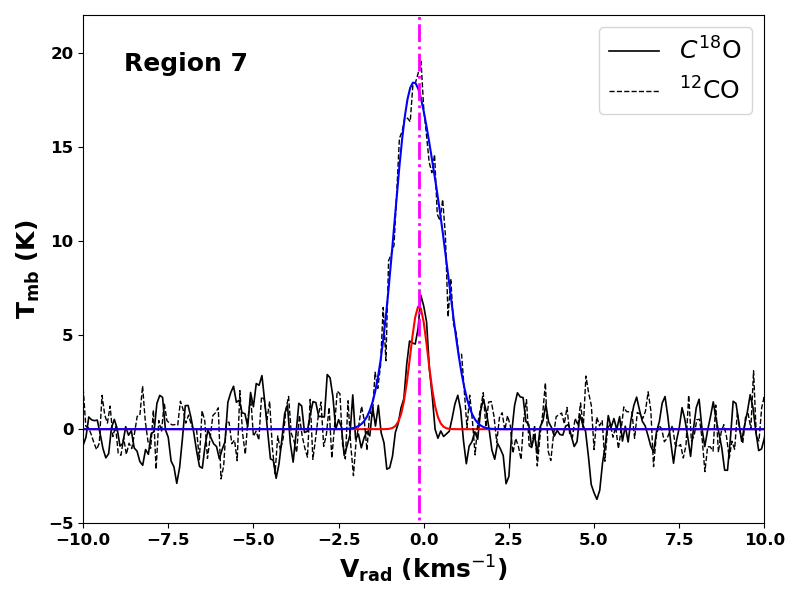}
    \end{subfigure}%
    \centering
    \begin{subfigure}[t]{0.35\textwidth}
        \centering
        \includegraphics[height=1.85in]{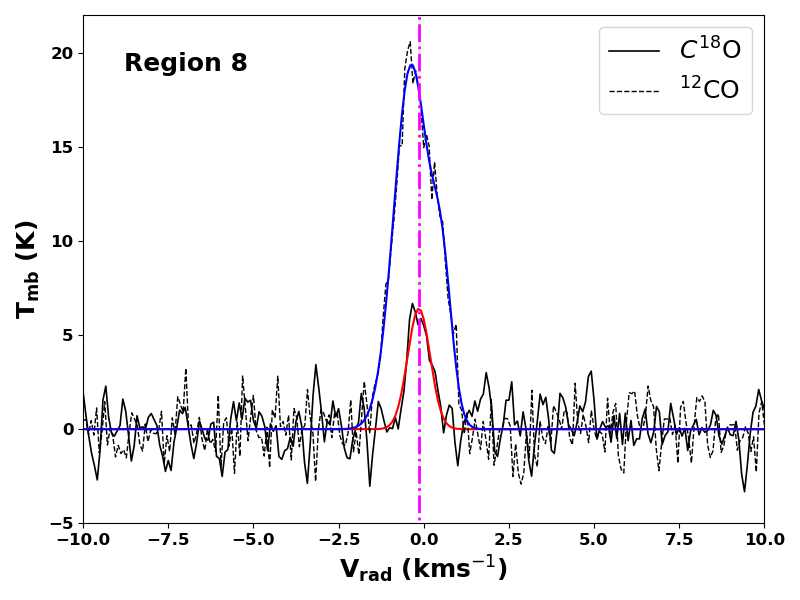}
    \end{subfigure}
    \begin{subfigure}[t]{0.35\textwidth}
        \centering
        \includegraphics[height=1.85in]{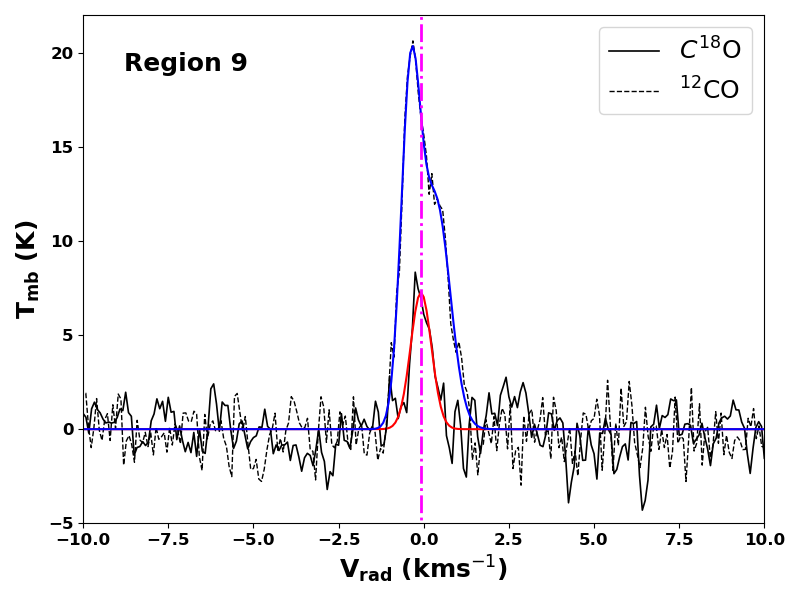}
    \end{subfigure}%
    \centering
    \begin{subfigure}[t]{0.35\textwidth}
        \centering
        \includegraphics[height=1.85in]{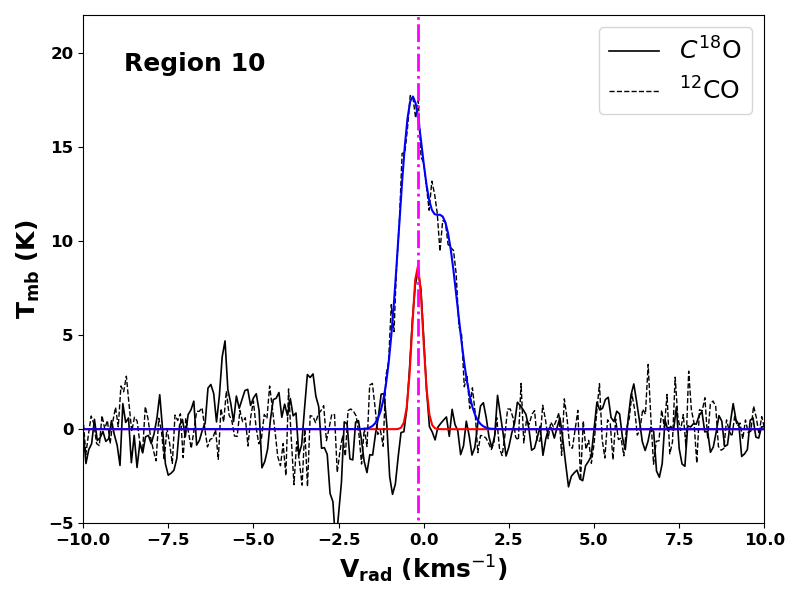}
    \end{subfigure}%
    \begin{subfigure}[t]{0.35\textwidth}
        \centering
        \includegraphics[height=1.85in]{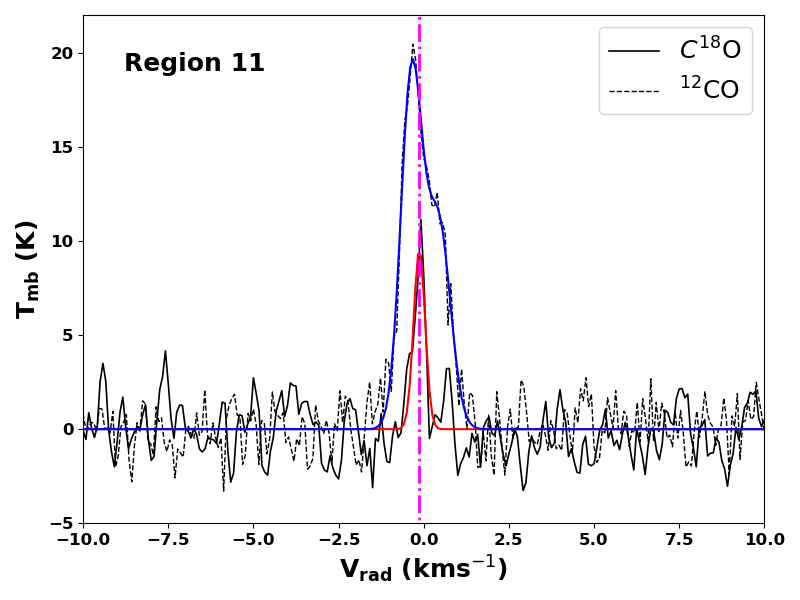}
    \end{subfigure}
    \centering
    \begin{subfigure}[t]{0.35\textwidth}
        \centering
        \includegraphics[height=1.85in]{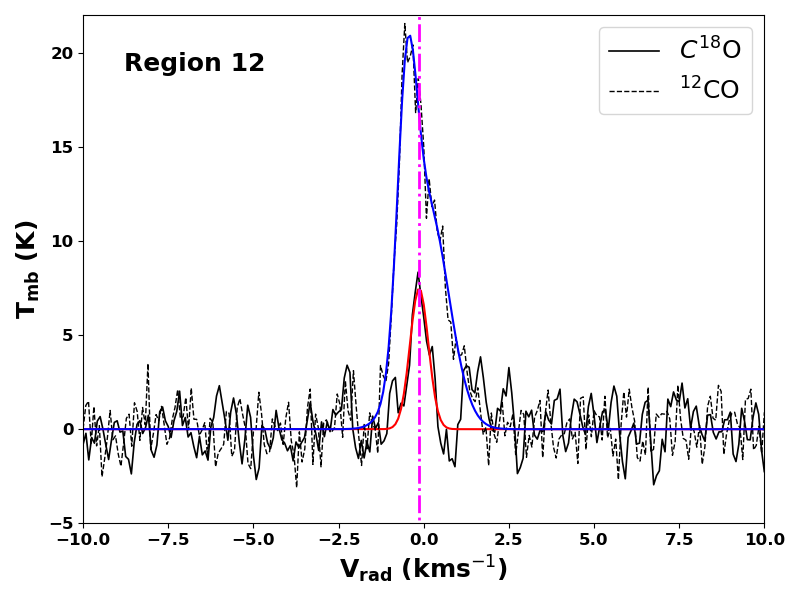}
    \end{subfigure}%
    \begin{subfigure}[t]{0.35\textwidth}
        \centering
        \includegraphics[height=1.85in]{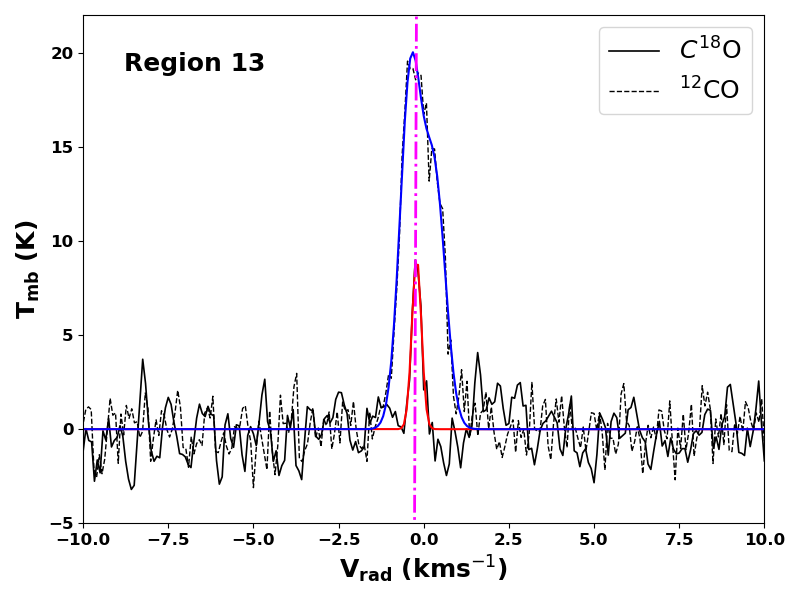}
    \end{subfigure}%
    \centering
    \begin{subfigure}[t]{0.35\textwidth}
        \centering
        \includegraphics[height=1.85in]{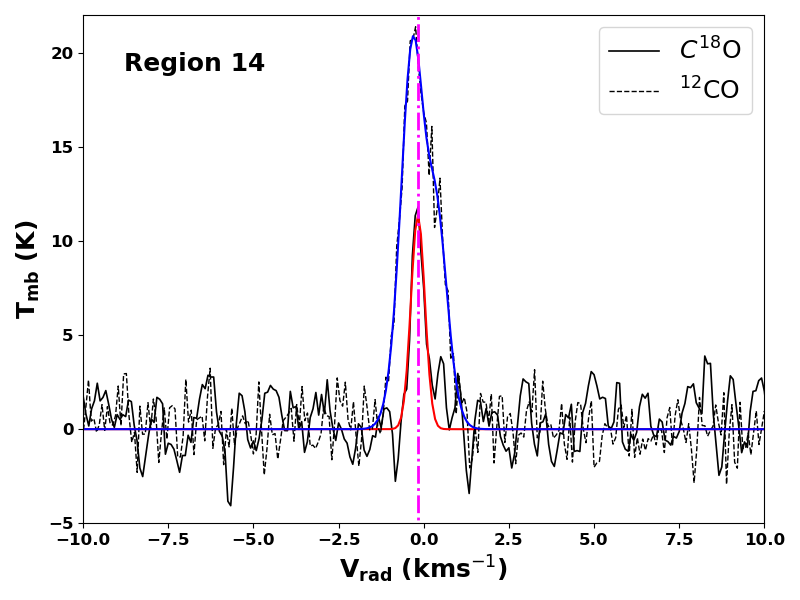}
    \end{subfigure}
    \begin{subfigure}[t]{0.35\textwidth}
        \centering
        \includegraphics[height=1.85in]{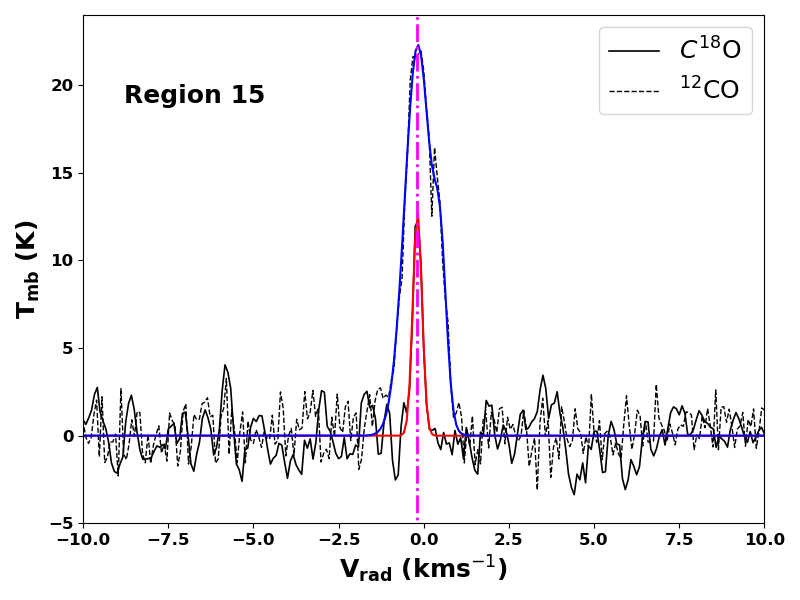}
    \end{subfigure}%
    \begin{subfigure}[t]{0.35\textwidth}
        \centering
        \includegraphics[height=1.85in]{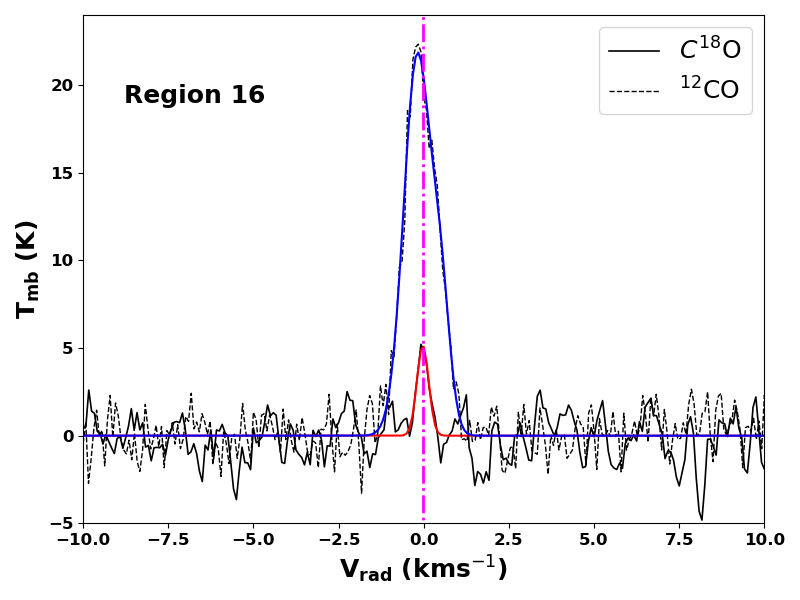}
    \end{subfigure}
     \caption{Spectra exhibiting a significant blue-skewed $^{12}$CO line profile (with $\delta V$ $<$ -0.25) are presented. The $^{12}$CO and C$^{18}$O emissions are averaged over regions marked by squares of size 45 arcsec (shown in Figure~\ref{fig:regions}). C$^{18}$O spectra are boxcar smoothed by 3 channels corresponding to a velocity smoothing of 0.24 km\,s$^{-1}$. The fit to  $^{12}$CO spectrum is depicted in blue and C$^{18}$O in red. The vertical, dashed magenta line corresponds to the LSR velocity estimated from the peak position of C$^{18}$O spectra.}
    \label{fig:Skewedprofile}
\end{figure*}

Since the $^{12}$CO spectra show blue asymmetry in their line profiles, it is essential to quantify the degree of observed blue-skewness. A robust quantity characterizing the degree of line asymmetry is the asymmetry parameter, $\delta V$. This dimensionless parameter is defined as the ratio of the difference between the peak velocities of the optically thick line ($V_{\rm thick}$) and the optically thin line ($V_{\rm thin}$) and the FWHM of the optically thin line (\(\Delta \)$V_{\rm thin}$). The criteria for a significant blue-skewed line profile is that the asymmetry parameter, $\delta V$ $<$ -0.25 \citep{1997ApJ...489..719M}.  We quantify the blue asymmetry by estimating  $\delta V$ in 16 regions where we visually observed a blue-skewed line profile. These regions of size 45 arcsec are overlaid on the $^{12}$CO\,(1\textendash\,0) channel map of the cometary head of LBN 437 with a velocity range of -0.238 to -0.159 km\,s$^{-1}$ as depicted in  Figure~\ref{fig:regions}.
Following \citet{1997ApJ...489..719M}, we estimate $\delta V$ using the equation, 
\begin{eqnarray*} 
   \delta V = \frac{(V_{\rm thick}-V_{\rm thin})}{\Delta V_{\rm thin}}.
\end{eqnarray*}
Out of the 16 regions considered, 11 regions exhibit $\delta V$ $<$ -0.25, thus showing a significant blue-skewed $^{12}$CO line profile. The $^{12}$CO spectra with a significant blue-skewed line profile ($\delta V$ $<$ -0.25) are depicted in Figure~\ref{fig:Skewedprofile} along with the C$^{18}$O spectra. We fit a double gaussian to the $^{12}$CO emission to account for the skewness observed. The double gaussian and single gaussian fit to the $^{12}$CO and C$^{18}$O line profiles are shown in blue and red, respectively. The LSR velocity estimated by identifying the peak positions of the C$^{18}$O line is also indicated using a vertical dashed line. The regions mentioned in the plots correspond to the regions marked on the $^{12}$CO\,(1\textendash\,0) channel map of the cometary head of LBN 437 as shown in Figure~\ref{fig:regions}.
\\To reinforce the idea of gas infall motion occurring in the cometary head of LBN 437, we determine the infall velocity in the regions characterized by an infall signature. Infall velocity ($V_{\rm inf}$) is defined as the difference in the peak velocities of the optically thin and thick line. It can be represented as $V_{\rm inf}$ = $V_{\rm thin}$ - $V_{\rm thick}$. We calculate $V_{\rm inf}$ only in the regions that show significant blue-skewness with $\delta V$ $<$ -0.25.  Including regions with $\delta V$ $>$ -0.25, might introduce uncertainties in the estimated $V_{\rm inf}$, as the skewness might be caused due to non-infall dynamics such as rotation \citep{2004MNRAS.352.1365R}. We estimate an average infall velocity of 0.25 km\,s$^{-1}$ in the cometary head of LBN 437. Table~\ref{tab:Asymmetryparameter} shows the asymmetry parameter and infall velocity estimated in regions of size 45 arcsec where the spectra of $^{12}$CO emission show a blue$-$skeweness. Here, the gaussian fit to the C$^{18}$O spectra yields $V_{\rm thin}$  and the peak velocity of the blue$-$shifted component of $^{12}$CO yields $V_{\rm thick}$. The velocity width of the C$^{18}$O line yields $\Delta V$$_{\rm thin}$.
\begin{table}
	\centering
	\caption{Estimated $^{12}$CO  Asymmetry Parameter ($\delta V$) and infall velocity ($V_{\rm inf}$ in km\,s$^{-1}$) in regions where the spectra are averaged over a size of 45 arcsec. $V_{\rm thick}$ and $V_{\rm thin}$ are the peak velocities of $^{12}$CO and C$^{18}$O lines in km\,s$^{-1}$ respectively and the FWHM of C$^{18}$O line in km\,s$^{-1}$ is \(\Delta \)$V_{\rm thin}$.} 
	\label{tab:Asymmetryparameter}
	\begin{tabular}{lccccr} 
		\hline
		Region & $V_{\rm thin}$ & $V_{\rm thick}$ & \(\Delta \)$V_{\rm thin}$ & $\delta V$ & $V_{\rm inf}$\\
		\hline
		1 & -0.22 & -0.14 & 0.54 & 0.15 & -\\
		2 & -0.13 & -0.24 & 0.49 & -0.22 & -\\
		3 & -0.3 & -0.21 & 0.8 & 0.11 &  -\\
		4 & -0.05 & -0.44 & 0.82 & -0.48 &  0.39 \\
		5 & -0.27 & -0.43 & 1.08 & -0.15 & -\\
		6 & -0.16 & -0.39 & 0.96 & -0.24 & -\\
		7 & -0.14 & -0.37 & 0.61 & -0.38 & 0.23\\
		8 & -0.14 & -0.37 & 0.78 & -0.29 & 0.23\\
		9 & -0.08 & -0.39 & 0.73 & -0.42 & 0.31\\
		10 & -0.18 & -0.36 & 0.38 & -0.47 & 0.23\\
		11 & -0.13 & -0.36 & 0.4 & -0.58 & 0.23\\
		12 & -0.13 & -0.5 & 0.63 & -0.59 & 0.37\\
		13 & -0.21 & -0.49 & 0.33 & -0.85 & 0.28\\
		14 & -0.17 & -0.31 & 0.47 & -0.3 & 0.14\\
		15 & -0.18 & -0.32 & 0.33 & -0.42 & 0.14\\
		16 & -0.03 & -0.19 & 0.4 & -0.4 & 0.16\\
		\hline
	\end{tabular}
\end{table}
\\
\\We determine the mass infall rate ($\dot{M}_{\rm inf}$) in the observed region using the equation given by \citet{2010A&A...517A..66L},
\begin{eqnarray*} 
  \dot{M}_{\rm inf}=4\pi R^2V_{\rm inf}\rho
\end{eqnarray*}
where $\rho$, the average volume density, is given by $\frac{M}{4/3 \pi R^3}$ and $V_{\rm inf}$ is the average infall velocity. We determined the mass infall rate by considering the average $V_{\rm inf}$ estimated from regions exhibiting significant blue-skewness as these regions are the best tracers of infall motion with minimum contamination from other non-infall dynamics. Although localized variations may exist, the averaged $V_{\rm inf}$ reasonably approximates the dominant infall features and thus serves as a valid estimate of the global mass infall rate. In addition, we also lack detectable C$^{18}$O emission in the other regions, which further limits our measurement of blue-skewness. We estimate the effective radius ($R$) using the equation, $R$ = ($A$/$\pi$)$^{0.5}$, where $A$ is the area enclosed by the 3$\sigma$ contour corresponding to the $^{12}$CO TRAO data. This method of estimating the effective radius simply assumes that the region is circularly symmetric. However, CGs are, by definition, tear-shaped and significantly elongated. This can cause systematic errors in the estimated effective radius. For instance, a spherical assumption can underestimate the extent along the major axis, resulting in an overestimated average density. A better way to estimate the effective radius would be to use moment analysis or fit an ellipse to the projected shape \citep{Rosolowsky_2006,Ge_2024}.\\
Furthermore, we also estimate the mass ($M$) from the \textit{Planck} 857 GHz emission following the relation given by \citet{Hildebrand},
 \begin{eqnarray*} 
  M = \frac{S_{\rm \nu} D d^{2}}{K_{\rm \nu}B_{\rm \nu}(T)}
\end{eqnarray*}
where $S_{\rm \nu}$ is the flux density obtained from the 857 GHz \textit{Planck} emission, D is the  dust$-$to$-$gas mass ratio (assumed as 0.01), d is the distance to the cloud (taken as 360 pc), $K_{\rm \nu}$ is the dust opacity (adopted as 1.85 cm$^{2}$g$^{-1}$ from \citealp{1994A&A...291..943O}), and $B_{\rm \nu}(T)$ is the Planck function estimated assuming the dust temperature to be 27\,K, as determined for LBN 437 by Saikhom et al. (2025, in prep.). The effective radius estimated for the region is 0.53 pc. The mass of the region is 352 M$_{\odot}$ and the mass infall rate calculated is 5.08 $\times$ 10$^{-4}$ M$_{\odot}$ yr$^{-1}$. Previous studies reported that the mass infall rate in an intermediate or high$-$mass star-forming regions is about 10$^{-4}$ M$_{\odot}$ yr$^{-1}$ to 10$^{-2}$ M$_{\odot}$ yr$^{-1}$ \citep[e.g.][]{Wu_2009,2013ApJ...775L...2L,10.1093/mnras/stv2801,Yang_2021,Yu_2022} while that associated with a low$-$mass star-forming region varies between 10$^{-6}$ M$_{\odot}$ yr$^{-1}$ to 10$^{-5}$ M$_{\odot}$ yr$^{-1}$ \citep{2013A&A...549A...5R, 2015MNRAS.450.1926H,Kim2021ApJ...910..112K}. This suggests that the observed region in the cometary head of LBN 437 is a possible site for forming high-mass stars. Hence, from the observation of infall tracers and estimated mass infall rate, it is evident that the cometary head of LBN 437 is actively accreting material from the ambient medium, thereby supporting ongoing star formation. In addition, the presence of LkH$\alpha$ sources in the cometary head suggests that star formation is taking place in a clustered environment. This scenario is consistent with the competitive accretion model of high-mass star formation wherein protostars accrete by drawing material from a large common gas reservoir \citep{2001MNRAS.323..785B, 2006MNRAS.370..488B, Tan_2014, 2021A&A...648A.100B, 2024MNRAS.533.1075Z}.\\
Our findings demonstrate that the cometary head of LBN 437 is gravitationally contracting. The molecular gas kinematics further shows evidence of gas infall in the cometary head, highlighting the potential formation of high-mass stars. 
\section{Summary}
\label{sec:Summary}
Previous studies on LBN 437 have extensively examined the impact of Lac OB1 stars on its origin, morphology, and magnetic field orientation. Nevertheless, there appears to be a dearth of literature that addresses the kinematic properties of the cloud. The main motivation for this work is to investigate the kinematics of molecular gas in the cometary head of LBN 437. We analysed the first rotational transition (1\textendash\,0) of the $^{12}$CO and C$^{18}$O molecular lines obtained using TRAO. The results are summarized below.
\begin{itemize}
\item{We examined the averaged spectra of $^{12}$CO and C$^{18}$O molecular line profiles. The region has $V_{\rm LSR}$ close to 0.0 km\,s$^{-1}$, indicating that the cloud is stationary relative to the stars in the immediate vicinity. We find that the averaged $^{12}$CO spectrum shows a slightly skewed profile, revealing the possibility of the cloud as contracting.}\\
\item{We examined the gas motion in the cometary head by plotting the spectral grid map of $^{12}$CO emission. We identified asymmetric signatures in the line profile of $^{12}$CO emission, in particular, the blue-skewed line profiles in the inner region of the cometary head, near the LkH$\alpha$ sources.}\\
\item{The presence of blue-skewed $^{12}$CO line profiles suggests that the gas infall motion is occurring in the cometary head of LBN 437. To support the idea of infall, we estimated the infall velocity and mass infall rates towards the cometary head of LBN 437 as 0.25 km\,s$^{-1}$ and  
\,5.08 $\times$ 10$^{-4}$ M$_{\odot}$ yr$^{-1}$, respectively. The estimated mass infall rate aligns with that observed in the high-mass star-forming regions, indicating that the cometary head of LBN 437 is a potential site of high-mass star formation.}\\
\item{The estimated mass infall rate indicates a collapse phase, with accretion rates high enough to facilitate the formation of high-mass stars.  In conjunction with the presence of LkHa sources, this implies that LBN 437 might be undergoing clustered star formation consistent with models of competitive accretion.}
\end{itemize}
However, we require additional molecular line observations towards LBN 437 to fully comprehend the kinematic structure of the cloud and the impact of the ionizing radiation from the Lac OB1 stars on the gas motion in LBN 437.

\section*{Acknowledgements}
We thank the referee for the constructive comments and suggestions that have significantly enhanced the quality of the paper.
Aardra S acknowledges the Department of Science and Technology (DST) for the INSPIRE Fellowship
(IF220281). Also, thank the Centre for Research, CHRIST
(Deemed to be University) and Board of Graduate Studies, Indian Institute of Astrophysics (IIA), for all their support during the course of this work. Aardra S and A. Soam also thank IIA VSP programme for providing internship opportunity and supporting this work. A. Soam is thankful to the TRAO staff for their continuous support during the radio observations. N.I. acknowledges the support by the China Postdoctoral Science Foundation through grant No. 2023M733624 and the Shanghai Postdoctoral Excellence Program through grant No. 2023682. C.W.L. acknowledges support from the Basic Science Research Program through the NRF funded by the Ministry of Education, Science and Technology (NRF- 2019R1A2C1010851) and by the Korea Astronomy and Space Science Institute grant funded by the Korea government (MSIT; project No. 2024-1-841-00). We acknowledge the use of Planck data from the European Space Agency mission. This research has made use of the SIMBAD database operated at CDS, Strasbourg, France, and NASA/IPAC Infrared Science Archive operated by the Jet Propulsion Laboratory, Caltech, under contract with NASA.
\section*{Data Availability}
The data supporting the findings of this study will be available from the corresponding author upon reasonable request.



\bibliographystyle{mnras}
\bibliography{example} 

\begin{thebibliography}{}
\makeatletter
\relax
\def\mn@urlcharsother{\let\do\@makeother \do\$\do\&\do\#\do\^\do\_\do\%\do\~}
\def\mn@doi{\begingroup\mn@urlcharsother \@ifnextchar [ {\mn@doi@} {\mn@doi@[]}}
\def\mn@doi@[#1]#2{\def\@tempa{#1}\ifx\@tempa\@empty \href {http://dx.doi.org/#2} {doi:#2}\else \href {http://dx.doi.org/#2} {#1}\fi \endgroup}
\def\mn@eprint#1#2{\mn@eprint@#1:#2::\@nil}
\def\mn@eprint@arXiv#1{\href {http://arxiv.org/abs/#1} {{\tt arXiv:#1}}}
\def\mn@eprint@dblp#1{\href {http://dblp.uni-trier.de/rec/bibtex/#1.xml} {dblp:#1}}
\def\mn@eprint@#1:#2:#3:#4\@nil{\def\@tempa {#1}\def\@tempb {#2}\def\@tempc {#3}\ifx \@tempc \@empty \let \@tempc \@tempb \let \@tempb \@tempa \fi \ifx \@tempb \@empty \def\@tempb {arXiv}\fi \@ifundefined {mn@eprint@\@tempb}{\@tempb:\@tempc}{\expandafter \expandafter \csname mn@eprint@\@tempb\endcsname \expandafter{\@tempc}}}

\bibitem[\protect\citeauthoryear{{Bally}}{{Bally}}{2016}]{doi:10.1146/annurev-astro-081915-023341}
{Bally} J.,  2016, \mn@doi [\araa] {10.1146/annurev-astro-081915-023341}, \href {https://ui.adsabs.harvard.edu/abs/2016ARA&A..54..491B} {54, 491}

\bibitem[\protect\citeauthoryear{{Beltr{\'a}n} et~al.,}{{Beltr{\'a}n} et~al.}{2021}]{2021A&A...648A.100B}
{Beltr{\'a}n} M.~T.,  et~al., 2021, \mn@doi [\aap] {10.1051/0004-6361/202040121}, \href {https://ui.adsabs.harvard.edu/abs/2021A&A...648A.100B} {648, A100}

\bibitem[\protect\citeauthoryear{{Bertoldi}}{{Bertoldi}}{1989}]{bertoldi1989photoevaporation}
{Bertoldi} F.,  1989, \mn@doi [\apj] {10.1086/168055}, \href {https://ui.adsabs.harvard.edu/abs/1989ApJ...346..735B} {346, 735}

\bibitem[\protect\citeauthoryear{{Beuther}, {Schilke}, {Sridharan}, {Menten}, {Walmsley}  \& {Wyrowski}}{{Beuther} et~al.}{2002}]{2002A&A...383..892B}
{Beuther} H.,  {Schilke} P.,  {Sridharan} T.~K.,  {Menten} K.~M.,  {Walmsley} C.~M.,   {Wyrowski} F.,  2002, \mn@doi [\aap] {10.1051/0004-6361:20011808}, \href {https://ui.adsabs.harvard.edu/abs/2002A&A...383..892B} {383, 892}

\bibitem[\protect\citeauthoryear{{Bonnell} \& {Bate}}{{Bonnell} \& {Bate}}{2006}]{2006MNRAS.370..488B}
{Bonnell} I.~A.,  {Bate} M.~R.,  2006, \mn@doi [\mnras] {10.1111/j.1365-2966.2006.10495.x}, \href {https://ui.adsabs.harvard.edu/abs/2006MNRAS.370..488B} {370, 488}

\bibitem[\protect\citeauthoryear{{Bonnell}, {Bate}, {Clarke}  \& {Pringle}}{{Bonnell} et~al.}{2001}]{2001MNRAS.323..785B}
{Bonnell} I.~A.,  {Bate} M.~R.,  {Clarke} C.~J.,   {Pringle} J.~E.,  2001, \mn@doi [\mnras] {10.1046/j.1365-8711.2001.04270.x}, \href {https://ui.adsabs.harvard.edu/abs/2001MNRAS.323..785B} {323, 785}

\bibitem[\protect\citeauthoryear{{Bourke}, {Hyland}, {Robinson}, {James}  \& {Wright}}{{Bourke} et~al.}{1995}]{bourke1995studies}
{Bourke} T.~L.,  {Hyland} A.~R.,  {Robinson} G.,  {James} S.~D.,   {Wright} C.~M.,  1995, \mn@doi [\mnras] {10.1093/mnras/276.4.1067}, \href {https://ui.adsabs.harvard.edu/abs/1995MNRAS.276.1067B} {276, 1067}

\bibitem[\protect\citeauthoryear{{Brand}, {Hawarden}, {Longmore}, {Williams}  \& {Caldwell}}{{Brand} et~al.}{1983}]{Brand}
{Brand} P.~W.~J.~L.,  {Hawarden} T.~G.,  {Longmore} A.~J.,  {Williams} P.~M.,   {Caldwell} J.~A.~R.,  1983, \mn@doi [\mnras] {10.1093/mnras/203.1.215}, \href {https://ui.adsabs.harvard.edu/abs/1983MNRAS.203..215B} {203, 215}

\bibitem[\protect\citeauthoryear{{Cerqueira}, {Cant{\'o}}, {Raga}  \& {Vasconcelos}}{{Cerqueira} et~al.}{2006}]{2006RMxAA..42..203C}
{Cerqueira} A.~H.,  {Cant{\'o}} J.,  {Raga} A.~C.,   {Vasconcelos} M.~J.,  2006, \mn@doi [\rmxaa] {10.48550/arXiv.astro-ph/0605625}, \href {https://ui.adsabs.harvard.edu/abs/2006RMxAA..42..203C} {42, 203}

\bibitem[\protect\citeauthoryear{{Chauhan}, {Pandey}, {Ogura}, {Jose}, {Ojha}, {Samal}  \& {Mito}}{{Chauhan} et~al.}{2011}]{2011MNRAS.415.1202C}
{Chauhan} N.,  {Pandey} A.~K.,  {Ogura} K.,  {Jose} J.,  {Ojha} D.~K.,  {Samal} M.~R.,   {Mito} H.,  2011, \mn@doi [\mnras] {10.1111/j.1365-2966.2011.18742.x}, \href {https://ui.adsabs.harvard.edu/abs/2011MNRAS.415.1202C} {415, 1202}

\bibitem[\protect\citeauthoryear{{Cortes-Rangel}, {Zapata}, {Toal{\'a}}, {Ho}, {Takahashi}, {Mesa-Delgado}  \& {Masqu{\'e}}}{{Cortes-Rangel} et~al.}{2020}]{Cortes-Rangel_2020}
{Cortes-Rangel} G.,  {Zapata} L.~A.,  {Toal{\'a}} J.~A.,  {Ho} P. T.~P.,  {Takahashi} S.,  {Mesa-Delgado} A.,   {Masqu{\'e}} J.~M.,  2020, \mn@doi [\aj] {10.3847/1538-3881/ab6295}, \href {https://ui.adsabs.harvard.edu/abs/2020AJ....159...62C} {159, 62}

\bibitem[\protect\citeauthoryear{{Dorschner} \& {G{\"u}rtler}}{{Dorschner} \& {G{\"u}rtler}}{1964}]{Dorschner}
{Dorschner} J.,  {G{\"u}rtler} J.,  1964, \mn@doi [Astronomische Nachrichten] {10.1002/asna.19652880106}, \href {https://ui.adsabs.harvard.edu/abs/1964AN....288...23D} {288, 23}

\bibitem[\protect\citeauthoryear{{Elmegreen} \& {Lada}}{{Elmegreen} \& {Lada}}{1977}]{1977ApJ...214..725E}
{Elmegreen} B.~G.,  {Lada} C.~J.,  1977, \mn@doi [\apj] {10.1086/155302}, \href {https://ui.adsabs.harvard.edu/abs/1977ApJ...214..725E} {214, 725}

\bibitem[\protect\citeauthoryear{{Evans}}{{Evans}}{1999}]{1999ARA&A..37..311E}
{Evans} Neal~J. I.,  1999, \mn@doi [\araa] {10.1146/annurev.astro.37.1.311}, \href {https://ui.adsabs.harvard.edu/abs/1999ARA&A..37..311E} {37, 311}

\bibitem[\protect\citeauthoryear{{Fuller}, {Williams}  \& {Sridharan}}{{Fuller} et~al.}{2005}]{2005A&A...442..949F}
{Fuller} G.~A.,  {Williams} S.~J.,   {Sridharan} T.~K.,  2005, \mn@doi [\aap] {10.1051/0004-6361:20042110}, \href {https://ui.adsabs.harvard.edu/abs/2005A&A...442..949F} {442, 949}

\bibitem[\protect\citeauthoryear{{Gao} \& {Lou}}{{Gao} \& {Lou}}{2010}]{Gao}
{Gao} Y.,  {Lou} Y.-Q.,  2010, \mn@doi [\mnras] {10.1111/j.1365-2966.2009.15651.x}, \href {https://ui.adsabs.harvard.edu/abs/2010MNRAS.403.1919G} {403, 1919}

\bibitem[\protect\citeauthoryear{{Garcia-Segura} \& {Franco}}{{Garcia-Segura} \& {Franco}}{1996}]{1996ApJ...469..171G}
{Garcia-Segura} G.,  {Franco} J.,  1996, \mn@doi [\apj] {10.1086/177769}, \href {https://ui.adsabs.harvard.edu/abs/1996ApJ...469..171G} {469, 171}

\bibitem[\protect\citeauthoryear{{Ge}, {Du}  \& {Yuan}}{{Ge} et~al.}{2024}]{Ge_2024}
{Ge} W.,  {Du} F.,   {Yuan} L.,  2024, \mn@doi [\mnras] {10.1093/mnras/stae680}, \href {https://ui.adsabs.harvard.edu/abs/2024MNRAS.529.3060G} {529, 3060}

\bibitem[\protect\citeauthoryear{{Haikala} \& {Olberg}}{{Haikala} \& {Olberg}}{2007}]{Haikala}
{Haikala} L.~K.,  {Olberg} M.,  2007, \mn@doi [\aap] {10.1051/0004-6361:20065256}, \href {https://ui.adsabs.harvard.edu/abs/2007A&A...466..191H} {466, 191}

\bibitem[\protect\citeauthoryear{{Hawarden} \& {Brand}}{{Hawarden} \& {Brand}}{1976}]{Hawarden}
{Hawarden} T.~G.,  {Brand} P.~W.~J.~L.,  1976, \mn@doi [\mnras] {10.1093/mnras/175.1.19P}, \href {https://ui.adsabs.harvard.edu/abs/1976MNRAS.175P..19H} {175, 19P}

\bibitem[\protect\citeauthoryear{{He} et~al.,}{{He} et~al.}{2015}]{2015MNRAS.450.1926H}
{He} Y.-X.,  et~al., 2015, \mn@doi [\mnras] {10.1093/mnras/stv732}, \href {https://ui.adsabs.harvard.edu/abs/2015MNRAS.450.1926H} {450, 1926}

\bibitem[\protect\citeauthoryear{{Hern{\'a}ndez}, {Calvet}, {Brice{\~n}o}, {Hartmann}  \& {Berlind}}{{Hern{\'a}ndez} et~al.}{2004}]{hernandez2004spectral}
{Hern{\'a}ndez} J.,  {Calvet} N.,  {Brice{\~n}o} C.,  {Hartmann} L.,   {Berlind} P.,  2004, \mn@doi [\aj] {10.1086/381908}, \href {https://ui.adsabs.harvard.edu/abs/2004AJ....127.1682H} {127, 1682}

\bibitem[\protect\citeauthoryear{{Hildebrand}}{{Hildebrand}}{1983}]{Hildebrand}
{Hildebrand} R.~H.,  1983, \qjras, \href {https://ui.adsabs.harvard.edu/abs/1983QJRAS..24..267H} {24, 267}

\bibitem[\protect\citeauthoryear{{Jackson} et~al.,}{{Jackson} et~al.}{2019}]{2019ApJ...870....5J}
{Jackson} J.~M.,  et~al., 2019, \mn@doi [\apj] {10.3847/1538-4357/aaef84}, \href {https://ui.adsabs.harvard.edu/abs/2019ApJ...870....5J} {870, 5}

\bibitem[\protect\citeauthoryear{{Kajdic}, {Reipurth}, {Raga}  \& {Walawender}}{{Kajdic} et~al.}{2010}]{2010RMxAA..46...67K}
{Kajdic} P.,  {Reipurth} B.,  {Raga} A.~C.,   {Walawender} J.,  2010, \mn@doi [\rmxaa] {10.48550/arXiv.1003.4327}, \href {https://ui.adsabs.harvard.edu/abs/2010RMxAA..46...67K} {46, 67}

\bibitem[\protect\citeauthoryear{Khavtassi}{Khavtassi}{1960}]{khavtassi1960atlas}
Khavtassi J.~S.,  1960, Obs., Tbilisi

\bibitem[\protect\citeauthoryear{{Kim}, {Lee}, {Maheswar}, {Myers}  \& {Kim}}{{Kim} et~al.}{2021}]{Kim2021ApJ...910..112K}
{Kim} M.-R.,  {Lee} C.~W.,  {Maheswar} G.,  {Myers} P.~C.,   {Kim} G.,  2021, \mn@doi [\apj] {10.3847/1538-4357/abe4d3}, \href {https://ui.adsabs.harvard.edu/abs/2021ApJ...910..112K} {910, 112}

\bibitem[\protect\citeauthoryear{{Kuznetsova}, {Hartmann}  \& {Ballesteros-Paredes}}{{Kuznetsova} et~al.}{2018}]{10.1093/mnras/stx2480}
{Kuznetsova} A.,  {Hartmann} L.,   {Ballesteros-Paredes} J.,  2018, \mn@doi [\mnras] {10.1093/mnras/stx2480}, \href {https://ui.adsabs.harvard.edu/abs/2018MNRAS.473.2372K} {473, 2372}

\bibitem[\protect\citeauthoryear{{Lee} \& {Chen}}{{Lee} \& {Chen}}{2007}]{leechen}
{Lee} H.-T.,  {Chen} W.~P.,  2007, \mn@doi [\apj] {10.1086/510893}, \href {https://ui.adsabs.harvard.edu/abs/2007ApJ...657..884L} {657, 884}

\bibitem[\protect\citeauthoryear{{Lee}, {Myers}  \& {Tafalla}}{{Lee} et~al.}{1999}]{1999ApJ...526..788L}
{Lee} C.~W.,  {Myers} P.~C.,   {Tafalla} M.,  1999, \mn@doi [\apj] {10.1086/308027}, \href {https://ui.adsabs.harvard.edu/abs/1999ApJ...526..788L} {526, 788}

\bibitem[\protect\citeauthoryear{{Lefloch} \& {Lazareff}}{{Lefloch} \& {Lazareff}}{1994}]{lefloch1994cometary}
{Lefloch} B.,  {Lazareff} B.,  1994, \aap, \href {https://ui.adsabs.harvard.edu/abs/1994A&A...289..559L} {289, 559}

\bibitem[\protect\citeauthoryear{{Leung} \& {Brown}}{{Leung} \& {Brown}}{1977}]{1977ApJ...214L..73L}
{Leung} C.~M.,  {Brown} R.~L.,  1977, \mn@doi [\apjl] {10.1086/182446}, \href {https://ui.adsabs.harvard.edu/abs/1977ApJ...214L..73L} {214, L73}

\bibitem[\protect\citeauthoryear{{Liu}, {Wu}  \& {Zhang}}{{Liu} et~al.}{2013}]{2013ApJ...775L...2L}
{Liu} T.,  {Wu} Y.,   {Zhang} H.,  2013, \mn@doi [\apjl] {10.1088/2041-8205/775/1/L2}, \href {https://ui.adsabs.harvard.edu/abs/2013ApJ...775L...2L} {775, L2}

\bibitem[\protect\citeauthoryear{{Liu} et~al.,}{{Liu} et~al.}{2018}]{2018ApJS..234...28L}
{Liu} T.,  et~al., 2018, \mn@doi [\apjs] {10.3847/1538-4365/aaa3dd}, \href {https://ui.adsabs.harvard.edu/abs/2018ApJS..234...28L} {234, 28}

\bibitem[\protect\citeauthoryear{{Lo}, {Wiles}, {Redman}, {Cunningham}, {Bains}, {Jones}, {Burton}  \& {Bronfman}}{{Lo} et~al.}{2015}]{2015MNRAS.453.3245L}
{Lo} N.,  {Wiles} B.,  {Redman} M.~P.,  {Cunningham} M.~R.,  {Bains} I.,  {Jones} P.~A.,  {Burton} M.~G.,   {Bronfman} L.,  2015, \mn@doi [\mnras] {10.1093/mnras/stv1880}, \href {https://ui.adsabs.harvard.edu/abs/2015MNRAS.453.3245L} {453, 3245}

\bibitem[\protect\citeauthoryear{{L{\'o}pez-Sepulcre}, {Cesaroni}  \& {Walmsley}}{{L{\'o}pez-Sepulcre} et~al.}{2010}]{2010A&A...517A..66L}
{L{\'o}pez-Sepulcre} A.,  {Cesaroni} R.,   {Walmsley} C.~M.,  2010, \mn@doi [\aap] {10.1051/0004-6361/201014252}, \href {https://ui.adsabs.harvard.edu/abs/2010A&A...517A..66L} {517, A66}

\bibitem[\protect\citeauthoryear{{Mackey} \& {Lim}}{{Mackey} \& {Lim}}{2010}]{2010MNRAS.403..714M}
{Mackey} J.,  {Lim} A.~J.,  2010, \mn@doi [\mnras] {10.1111/j.1365-2966.2009.16181.x}, \href {https://ui.adsabs.harvard.edu/abs/2010MNRAS.403..714M} {403, 714}

\bibitem[\protect\citeauthoryear{{Maheswar} \& {Bhatt}}{{Maheswar} \& {Bhatt}}{2008}]{2008Ap&SS.315..215M}
{Maheswar} G.,  {Bhatt} H.~C.,  2008, \mn@doi [\apss] {10.1007/s10509-008-9822-7}, \href {https://ui.adsabs.harvard.edu/abs/2008Ap&SS.315..215M} {315, 215}

\bibitem[\protect\citeauthoryear{{Maheswar}, {Sharma}, {Biman}, {Pandey}  \& {Bhatt}}{{Maheswar} et~al.}{2007}]{Maheswar}
{Maheswar} G.,  {Sharma} S.,  {Biman} J.~M.,  {Pandey} A.~K.,   {Bhatt} H.~C.,  2007, \mn@doi [\mnras] {10.1111/j.1365-2966.2007.12020.x}, \href {https://ui.adsabs.harvard.edu/abs/2007MNRAS.379.1237M} {379, 1237}

\bibitem[\protect\citeauthoryear{{M{\"a}kel{\"a}} \& {Haikala}}{{M{\"a}kel{\"a}} \& {Haikala}}{2013}]{Makela}
{M{\"a}kel{\"a}} M.~M.,  {Haikala} L.~K.,  2013, \mn@doi [\aap] {10.1051/0004-6361/201220027}, \href {https://ui.adsabs.harvard.edu/abs/2013A&A...550A..83M} {550, A83}

\bibitem[\protect\citeauthoryear{{Mardones}, {Myers}, {Tafalla}, {Wilner}, {Bachiller}  \& {Garay}}{{Mardones} et~al.}{1997}]{1997ApJ...489..719M}
{Mardones} D.,  {Myers} P.~C.,  {Tafalla} M.,  {Wilner} D.~J.,  {Bachiller} R.,   {Garay} G.,  1997, \mn@doi [\apj] {10.1086/304812}, \href {https://ui.adsabs.harvard.edu/abs/1997ApJ...489..719M} {489, 719}

\bibitem[\protect\citeauthoryear{{McKee} \& {Tan}}{{McKee} \& {Tan}}{2003}]{2003ApJ...585..850M}
{McKee} C.~F.,  {Tan} J.~C.,  2003, \mn@doi [\apj] {10.1086/346149}, \href {https://ui.adsabs.harvard.edu/abs/2003ApJ...585..850M} {585, 850}

\bibitem[\protect\citeauthoryear{{Mookerjea} \& {Sandell}}{{Mookerjea} \& {Sandell}}{2009}]{mookerjea2009star}
{Mookerjea} B.,  {Sandell} G.,  2009, \mn@doi [\apj] {10.1088/0004-637X/706/1/896}, \href {https://ui.adsabs.harvard.edu/abs/2009ApJ...706..896M} {706, 896}

\bibitem[\protect\citeauthoryear{{Motte}, {Bontemps}  \& {Louvet}}{{Motte} et~al.}{2018}]{2018ARA&A..56...41M}
{Motte} F.,  {Bontemps} S.,   {Louvet} F.,  2018, \mn@doi [\araa] {10.1146/annurev-astro-091916-055235}, \href {https://ui.adsabs.harvard.edu/abs/2018ARA&A..56...41M} {56, 41}

\bibitem[\protect\citeauthoryear{{Myers}, {Mardones}, {Tafalla}, {Williams}  \& {Wilner}}{{Myers} et~al.}{1996}]{1996ApJ...465L.133M}
{Myers} P.~C.,  {Mardones} D.,  {Tafalla} M.,  {Williams} J.~P.,   {Wilner} D.~J.,  1996, \mn@doi [\apjl] {10.1086/310146}, \href {https://ui.adsabs.harvard.edu/abs/1996ApJ...465L.133M} {465, L133}

\bibitem[\protect\citeauthoryear{{Olano}, {Walmsley}  \& {Wilson}}{{Olano} et~al.}{1994}]{olano1994molecular}
{Olano} C.~A.,  {Walmsley} C.~M.,   {Wilson} T.~L.,  1994, \aap, \href {https://ui.adsabs.harvard.edu/abs/1994A&A...290..235O} {290, 235}

\bibitem[\protect\citeauthoryear{{Ortega}, {Paron}, {Giacani}, {Rubio}  \& {Dubner}}{{Ortega} et~al.}{2013}]{2013A&A...556A.105O}
{Ortega} M.~E.,  {Paron} S.,  {Giacani} E.,  {Rubio} M.,   {Dubner} G.,  2013, \mn@doi [\aap] {10.1051/0004-6361/201321808}, \href {https://ui.adsabs.harvard.edu/abs/2013A&A...556A.105O} {556, A105}

\bibitem[\protect\citeauthoryear{{Ossenkopf} \& {Henning}}{{Ossenkopf} \& {Henning}}{1994}]{1994A&A...291..943O}
{Ossenkopf} V.,  {Henning} T.,  1994, \aap, \href {https://ui.adsabs.harvard.edu/abs/1994A&A...291..943O} {291, 943}

\bibitem[\protect\citeauthoryear{{Peretto} et~al.,}{{Peretto} et~al.}{2013}]{2013A&A...555A.112P}
{Peretto} N.,  et~al., 2013, \mn@doi [\aap] {10.1051/0004-6361/201321318}, \href {https://ui.adsabs.harvard.edu/abs/2013A&A...555A.112P} {555, A112}

\bibitem[\protect\citeauthoryear{Peñaloza, Clark, Glover, Shetty  \& Klessen}{Peñaloza et~al.}{2016}]{Pe_aloza_2016}
Peñaloza C.~H.,  Clark P.~C.,  Glover S. C.~O.,  Shetty R.,   Klessen R.~S.,  2016, \mn@doi [\mnras] {10.1093/mnras/stw2892}, 465, 2277

\bibitem[\protect\citeauthoryear{{Planck Collaboration} et~al.,}{{Planck Collaboration} et~al.}{2014}]{2014A&A...571A...8P}
{Planck Collaboration} et~al., 2014, \mn@doi [\aap] {10.1051/0004-6361/201321538}, \href {https://ui.adsabs.harvard.edu/abs/2014A&A...571A...8P} {571, A8}

\bibitem[\protect\citeauthoryear{{Planck Collaboration} et~al.,}{{Planck Collaboration} et~al.}{2016a}]{2016A&A...586A.138P}
{Planck Collaboration} et~al., 2016a, \mn@doi [\aap] {10.1051/0004-6361/201525896}, \href {https://ui.adsabs.harvard.edu/abs/2016A&A...586A.138P} {586, A138}

\bibitem[\protect\citeauthoryear{{Planck Collaboration} et~al.,}{{Planck Collaboration} et~al.}{2016b}]{2016A&A...594A...8P}
{Planck Collaboration} et~al., 2016b, \mn@doi [\aap] {10.1051/0004-6361/201525820}, \href {https://ui.adsabs.harvard.edu/abs/2016A&A...594A...8P} {594, A8}

\bibitem[\protect\citeauthoryear{Qin, Schilke, Wu, Liu, Wu, S{\'a}nchez-Monge  \& Liu}{Qin et~al.}{2015}]{10.1093/mnras/stv2801}
Qin S.-L.,  Schilke P.,  Wu J.,  Liu T.,  Wu Y.,  S{\'a}nchez-Monge {\'A}.,   Liu Y.,  2015, \mn@doi [\mnras] {10.1093/mnras/stv2801}, 456, 2681

\bibitem[\protect\citeauthoryear{{Rathborne} et~al.,}{{Rathborne} et~al.}{2014}]{2014ApJ...786..140R}
{Rathborne} J.~M.,  et~al., 2014, \mn@doi [\apj] {10.1088/0004-637X/786/2/140}, \href {https://ui.adsabs.harvard.edu/abs/2014ApJ...786..140R} {786, 140}

\bibitem[\protect\citeauthoryear{{Rawat} et~al.,}{{Rawat} et~al.}{2024}]{2024MNRAS.528.2199R}
{Rawat} V.,  et~al., 2024, \mn@doi [\mnras] {10.1093/mnras/stae060}, \href {https://ui.adsabs.harvard.edu/abs/2024MNRAS.528.2199R} {528, 2199}

\bibitem[\protect\citeauthoryear{{Redman}, {Keto}, {Rawlings}  \& {Williams}}{{Redman} et~al.}{2004}]{2004MNRAS.352.1365R}
{Redman} M.~P.,  {Keto} E.,  {Rawlings} J.~M.~C.,   {Williams} D.~A.,  2004, \mn@doi [\mnras] {10.1111/j.1365-2966.2004.08027.x}, \href {https://ui.adsabs.harvard.edu/abs/2004MNRAS.352.1365R} {352, 1365}

\bibitem[\protect\citeauthoryear{{Reipurth}}{{Reipurth}}{1983}]{reipurth1983star}
{Reipurth} B.,  1983, \aap, \href {https://ui.adsabs.harvard.edu/abs/1983A&A...117..183R} {117, 183}

\bibitem[\protect\citeauthoryear{{Rosolowsky} \& {Leroy}}{{Rosolowsky} \& {Leroy}}{2006}]{Rosolowsky_2006}
{Rosolowsky} E.,  {Leroy} A.,  2006, \mn@doi [\pasp] {10.1086/502982}, \href {https://ui.adsabs.harvard.edu/abs/2006PASP..118..590R} {118, 590}

\bibitem[\protect\citeauthoryear{{Rygl}, {Wyrowski}, {Schuller}  \& {Menten}}{{Rygl} et~al.}{2013}]{2013A&A...549A...5R}
{Rygl} K.~L.~J.,  {Wyrowski} F.,  {Schuller} F.,   {Menten} K.~M.,  2013, \mn@doi [\aap] {10.1051/0004-6361/201219574}, \href {https://ui.adsabs.harvard.edu/abs/2013A&A...549A...5R} {549, A5}

\bibitem[\protect\citeauthoryear{{Saha}, {Soam}, {Baug}, {Gopinathan}, {Mondal}  \& {Ghosh}}{{Saha} et~al.}{2022}]{Saha_2022}
{Saha} P.,  {Soam} A.,  {Baug} T.,  {Gopinathan} M.,  {Mondal} S.,   {Ghosh} T.,  2022, \mn@doi [\mnras] {10.1093/mnras/stac943}, \href {https://ui.adsabs.harvard.edu/abs/2022MNRAS.513.2039S} {513, 2039}

\bibitem[\protect\citeauthoryear{{Sandford}, {Whitaker}  \& {Klein}}{{Sandford} et~al.}{1982}]{sandford1982radiation}
{Sandford} II M.~T.,  {Whitaker} R.~W.,   {Klein} R.~I.,  1982, \mn@doi [\apj] {10.1086/160245}, \href {https://ui.adsabs.harvard.edu/abs/1982ApJ...260..183S} {260, 183}

\bibitem[\protect\citeauthoryear{{Sandqvist}}{{Sandqvist}}{1976}]{Sandqvist}
{Sandqvist} A.,  1976, \mn@doi [\mnras] {10.1093/mnras/177.1.69P}, \href {https://ui.adsabs.harvard.edu/abs/1976MNRAS.177P..69S} {177, 69P}

\bibitem[\protect\citeauthoryear{{Schneider, N.} et~al.,}{{Schneider, N.} et~al.}{2015}]{Schneider_2015}
{Schneider, N.} et~al., 2015, \mn@doi [A&A] {10.1051/0004-6361/201424375}, 578, A29

\bibitem[\protect\citeauthoryear{{Sharma}, {Chen}, {Panwar}, {Sun}  \& {Gao}}{{Sharma} et~al.}{2022}]{2022ApJ...928...17S}
{Sharma} T.,  {Chen} W.~P.,  {Panwar} N.,  {Sun} Y.,   {Gao} Y.,  2022, \mn@doi [\apj] {10.3847/1538-4357/ac510b}, \href {https://ui.adsabs.harvard.edu/abs/2022ApJ...928...17S} {928, 17}

\bibitem[\protect\citeauthoryear{{Sharpless}}{{Sharpless}}{1959}]{sharpless}
{Sharpless} S.,  1959, \mn@doi [\apjs] {10.1086/190049}, \href {https://ui.adsabs.harvard.edu/abs/1959ApJS....4..257S} {4, 257}

\bibitem[\protect\citeauthoryear{Smith, Shetty, Stutz  \& Klessen}{Smith et~al.}{2012}]{Smith_2012}
Smith R.~J.,  Shetty R.,  Stutz A.~M.,   Klessen R.~S.,  2012, \mn@doi [\apj] {10.1088/0004-637X/750/1/64}, 750, 64

\bibitem[\protect\citeauthoryear{{Snell} \& {Loren}}{{Snell} \& {Loren}}{1977}]{1977ApJ...211..122S}
{Snell} R.~L.,  {Loren} R.~B.,  1977, \mn@doi [\apj] {10.1086/154909}, \href {https://ui.adsabs.harvard.edu/abs/1977ApJ...211..122S} {211, 122}

\bibitem[\protect\citeauthoryear{{Soam}}{{Soam}}{2021}]{2021RAA....21...87S}
{Soam} A.,  2021, \mn@doi [Research in Astronomy and Astrophysics] {10.1088/1674-4527/21/4/87}, \href {https://ui.adsabs.harvard.edu/abs/2021RAA....21...87S} {21, 087}

\bibitem[\protect\citeauthoryear{{Soam}, {Maheswar}, {Bhatt}, {Lee}  \& {Ramaprakash}}{{Soam} et~al.}{2013}]{soam2013magnetic}
{Soam} A.,  {Maheswar} G.,  {Bhatt} H.~C.,  {Lee} C.~W.,   {Ramaprakash} A.~N.,  2013, \mn@doi [\mnras] {10.1093/mnras/stt576}, \href {https://ui.adsabs.harvard.edu/abs/2013MNRAS.432.1502S} {432, 1502}

\bibitem[\protect\citeauthoryear{{Sugitani}, {Fukui}  \& {Ogura}}{{Sugitani} et~al.}{1991}]{Sugitani}
{Sugitani} K.,  {Fukui} Y.,   {Ogura} K.,  1991, \mn@doi [\apjs] {10.1086/191597}, \href {https://ui.adsabs.harvard.edu/abs/1991ApJS...77...59S} {77, 59}

\bibitem[\protect\citeauthoryear{Tan, Beltrán, Caselli, Fontani, Fuente, Krumholz, McKee  \& Stolte}{Tan et~al.}{2014}]{Tan_2014}
Tan J.~C.,  Beltrán M.~T.,  Caselli P.,  Fontani F.,  Fuente A.,  Krumholz M.~R.,  McKee C.~F.,   Stolte A.,  2014, Massive Star Formation.
University of Arizona Press, \mn@doi{10.2458/azu_uapress_9780816531240-ch007}

\bibitem[\protect\citeauthoryear{{Tremblin}, {Audit}, {Minier}, {Schmidt}  \& {Schneider}}{{Tremblin} et~al.}{2012}]{2012A&A...546A..33T}
{Tremblin} P.,  {Audit} E.,  {Minier} V.,  {Schmidt} W.,   {Schneider} N.,  2012, \mn@doi [\aap] {10.1051/0004-6361/201219224}, \href {https://ui.adsabs.harvard.edu/abs/2012A&A...546A..33T} {546, A33}

\bibitem[\protect\citeauthoryear{{Urquhart}, {Morgan}  \& {Thompson}}{{Urquhart} et~al.}{2009}]{2009A&A...497..789U}
{Urquhart} J.~S.,  {Morgan} L.~K.,   {Thompson} M.~A.,  2009, \mn@doi [\aap] {10.1051/0004-6361/200811149}, \href {https://ui.adsabs.harvard.edu/abs/2009A&A...497..789U} {497, 789}

\bibitem[\protect\citeauthoryear{{Vilas-Boas}, {Myers}  \& {Fuller}}{{Vilas-Boas} et~al.}{1994}]{Vilas-Boas}
{Vilas-Boas} J.~W.~S.,  {Myers} P.~C.,   {Fuller} G.~A.,  1994, \mn@doi [\apj] {10.1086/174628}, \href {https://ui.adsabs.harvard.edu/abs/1994ApJ...433...96V} {433, 96}

\bibitem[\protect\citeauthoryear{{Walch}, {Whitworth}, {Bisbas}, {W{\"u}nsch}  \& {Hubber}}{{Walch} et~al.}{2012}]{2012MNRAS.427..625W}
{Walch} S.~K.,  {Whitworth} A.~P.,  {Bisbas} T.,  {W{\"u}nsch} R.,   {Hubber} D.,  2012, \mn@doi [\mnras] {10.1111/j.1365-2966.2012.21767.x}, \href {https://ui.adsabs.harvard.edu/abs/2012MNRAS.427..625W} {427, 625}

\bibitem[\protect\citeauthoryear{{Walch}, {Whitworth}, {Bisbas}, {W{\"u}nsch}  \& {Hubber}}{{Walch} et~al.}{2013}]{walch2013clumps}
{Walch} S.,  {Whitworth} A.~P.,  {Bisbas} T.~G.,  {W{\"u}nsch} R.,   {Hubber} D.~A.,  2013, \mn@doi [\mnras] {10.1093/mnras/stt1115}, \href {https://ui.adsabs.harvard.edu/abs/2013MNRAS.435..917W} {435, 917}

\bibitem[\protect\citeauthoryear{{Williams}}{{Williams}}{1999}]{williams1999shadowing}
{Williams} R.~J.~R.,  1999, \mn@doi [\mnras] {10.1046/j.1365-8711.1999.03014.x}, \href {https://ui.adsabs.harvard.edu/abs/1999MNRAS.310..789W} {310, 789}

\bibitem[\protect\citeauthoryear{{Williams}, {Brand}, {Longmore}  \& {Hawarden}}{{Williams} et~al.}{1977}]{Williams}
{Williams} P.~M.,  {Brand} P.~W.~J.~L.,  {Longmore} A.~J.,   {Hawarden} T.~G.,  1977, \mn@doi [\mnras] {10.1093/mnras/180.4.709}, \href {https://ui.adsabs.harvard.edu/abs/1977MNRAS.180..709W} {180, 709}

\bibitem[\protect\citeauthoryear{{Wright} et~al.,}{{Wright} et~al.}{2010}]{Wright_2010}
{Wright} E.~L.,  et~al., 2010, \mn@doi [\aj] {10.1088/0004-6256/140/6/1868}, \href {https://ui.adsabs.harvard.edu/abs/2010AJ....140.1868W} {140, 1868}

\bibitem[\protect\citeauthoryear{{Wu}, {Qin}, {Guan}, {Xue}, {Ren}, {Liu}, {Huang}  \& {Chen}}{{Wu} et~al.}{2009}]{Wu_2009}
{Wu} Y.,  {Qin} S.-L.,  {Guan} X.,  {Xue} R.,  {Ren} Z.,  {Liu} T.,  {Huang} M.,   {Chen} S.,  2009, \mn@doi [\apjl] {10.1088/0004-637X/697/2/L116}, \href {https://ui.adsabs.harvard.edu/abs/2009ApJ...697L.116W} {697, L116}

\bibitem[\protect\citeauthoryear{{Yang}, {Jiang}, {Chen}, {Ao}  \& {Yu}}{{Yang} et~al.}{2021}]{Yang_2021}
{Yang} Y.,  {Jiang} Z.,  {Chen} Z.,  {Ao} Y.,   {Yu} S.,  2021, \mn@doi [\apj] {10.3847/1538-4357/ac22ab}, \href {https://ui.adsabs.harvard.edu/abs/2021ApJ...922..144Y} {922, 144}

\bibitem[\protect\citeauthoryear{{Yorke} \& {Sonnhalter}}{{Yorke} \& {Sonnhalter}}{2002}]{2002ApJ...569..846Y}
{Yorke} H.~W.,  {Sonnhalter} C.,  2002, \mn@doi [\apj] {10.1086/339264}, \href {https://ui.adsabs.harvard.edu/abs/2002ApJ...569..846Y} {569, 846}

\bibitem[\protect\citeauthoryear{Yu, Jiang, Yang, Chen  \& Feng}{Yu et~al.}{2022}]{Yu_2022}
Yu S.,  Jiang Z.,  Yang Y.,  Chen Z.,   Feng H.,  2022, \mn@doi [Research in Astronomy and Astrophysics] {10.1088/1674-4527/ac7d9d}, 22, 095014

\bibitem[\protect\citeauthoryear{{Zealey}, {Ninkov}, {Rice}, {Hartley}  \& {Tritton}}{{Zealey} et~al.}{1983}]{Zealey}
{Zealey} W.~J.,  {Ninkov} Z.,  {Rice} E.,  {Hartley} M.,   {Tritton} S.~B.,  1983, \aplett, \href {https://ui.adsabs.harvard.edu/abs/1983ApL....23..119Z} {23, 119}

\bibitem[\protect\citeauthoryear{{Zhang}, {Hunter}, {Brand}, {Sridharan}, {Molinari}, {Kramer}  \& {Cesaroni}}{{Zhang} et~al.}{2001}]{2001ApJ...552L.167Z}
{Zhang} Q.,  {Hunter} T.~R.,  {Brand} J.,  {Sridharan} T.~K.,  {Molinari} S.,  {Kramer} M.~A.,   {Cesaroni} R.,  2001, \mn@doi [\apjl] {10.1086/320345}, \href {https://ui.adsabs.harvard.edu/abs/2001ApJ...552L.167Z} {552, L167}

\bibitem[\protect\citeauthoryear{{Zhang}, {Cyganowski}, {Henshaw}, {Brogan}, {Hunter}, {Friesen}, {Bonnell}  \& {Viti}}{{Zhang} et~al.}{2024}]{2024MNRAS.533.1075Z}
{Zhang} S.,  {Cyganowski} C.~J.,  {Henshaw} J.~D.,  {Brogan} C.~L.,  {Hunter} T.~R.,  {Friesen} R.~K.,  {Bonnell} I.~A.,   {Viti} S.,  2024, \mn@doi [\mnras] {10.1093/mnras/stae1844}, \href {https://ui.adsabs.harvard.edu/abs/2024MNRAS.533.1075Z} {533, 1075}

\bibitem[\protect\citeauthoryear{{Zhou} \& {Evans}}{{Zhou} \& {Evans}}{1994}]{1994ASPC...65..183Z}
{Zhou} S.,  {Evans} N.~J. I.,  1994, in {Clemens} D.~P.,  {Barvainis} R.,  eds,  Astronomical Society of the Pacific Conference Series Vol. 65, Clouds, Cores, and Low Mass Stars. p.~183

\bibitem[\protect\citeauthoryear{{Zhou}, {Evans}, {Koempe}  \& {Walmsley}}{{Zhou} et~al.}{1993}]{1993ApJ...404..232Z}
{Zhou} S.,  {Evans} II N.~J.,  {Koempe} C.,   {Walmsley} C.~M.,  1993, \mn@doi [\apj] {10.1086/172271}, \href {https://ui.adsabs.harvard.edu/abs/1993ApJ...404..232Z} {404, 232}

\bibitem[\protect\citeauthoryear{{Zhou} et~al.,}{{Zhou} et~al.}{2021}]{10.1093/mnras/stab2801}
{Zhou} J.-W.,  et~al., 2021, \mn@doi [\mnras] {10.1093/mnras/stab2801}, \href {https://ui.adsabs.harvard.edu/abs/2021MNRAS.508.4639Z} {508, 4639}

\makeatother
\end{thebibliography}








\bsp	
\label{lastpage}
\end{document}